\documentclass[sigconf]{acmart}

\AtBeginDocument{%
  }

\setcopyright{none}

\usepackage{algorithm}
\usepackage[noend]{algpseudocode}
\usepackage{xspace}
\usepackage{threeparttable}
\usepackage{color}
\usepackage{amsfonts}
\usepackage{amsfonts}
\usepackage{wrapfig}
\usepackage{multirow}
\usepackage{listings}
\usepackage{comment}
\usepackage{booktabs}
\usepackage{makecell}
\usepackage{boxedminipage}
\usepackage{amsmath}
\usepackage{graphicx}
\usepackage{minibox}
\usepackage{pifont}
\usepackage{color}
\usepackage{xcolor}
\usepackage{subfiles}
\usepackage[utf8]{inputenc}
\usepackage[english]{babel}
\usepackage{xcolor}
\usepackage{ulem}
\usepackage{adjustbox}
\usepackage{xurl}
\usepackage{tcolorbox}
\newcommand{\bheading}[1]{{\vspace{3pt}\noindent{\textbf{#1}}}}
\newcolumntype{?}{!{\vrule width 1pt}}
\newcommand{\iheading}[1]{{\vspace{2pt}\noindent{\textit{#1}}}}

\newcounter{note}[section]

\colorlet{Mycolor1}{green!10!orange!90!}

\newcommand{\secref}[1]{\mbox{Sec.~\ref{#1}}\xspace}

\newcommand{\figref}[1]{\mbox{Fig.~\ref{#1}}}
\newcommand{\tabref}[1]{\mbox{Table~\ref{#1}}}

\newcommand{\ignore}[1]{}

\newcommand{\ie}{\textit{i.e.}\xspace}
\newcommand{\eg}{\textit{e.g.}\xspace}

\newcommand{\sysname}{\textsc{CachePrune}\xspace}

\newcounter{packednmbr}

\newenvironment{packeditemize}{
\begin{list}{$\bullet$}{
\setlength{\labelwidth}{0pt}
\setlength{\itemsep}{2pt}
\setlength{\leftmargin}{\labelwidth}
\addtolength{\leftmargin}{\labelsep}
\setlength{\parindent}{0pt}
\setlength{\listparindent}{\parindent}
\setlength{\parsep}{1pt}
\setlength{\topsep}{1pt}}}{\end{list}}

\newcounter{lessoncount}

\newtcolorbox{highlightbox}[1][]{
  colback=blue!5!white,
  colframe=blue!75!black,
  boxrule=0.8pt,
  arc=4pt,
  left=6pt,
  right=6pt,
  top=6pt,
  bottom=6pt,
  title=#1
}

\newtcolorbox{importantbox}[1][]{
  colback=red!5!white,
  colframe=red!75!black,
  boxrule=0.8pt,
  arc=4pt,
  left=6pt,
  right=6pt,
  top=6pt,
  bottom=6pt,
  title=#1
}

\newtcolorbox{graybox}[1][]{
  colback=gray!10,
  colframe=black!40,
  boxrule=0.5pt,
  arc=2pt,
  left=5pt,
  right=5pt,
  top=5pt,
  bottom=5pt,
  title=#1
}

\newtcolorbox{nobox}[1][]{
  colback=yellow!15,
  colframe=yellow!15,
  boxrule=0pt,
  arc=3pt,
  left=5pt,
  right=5pt,
  top=5pt,
  bottom=5pt,
  title=#1
}

\definecolor{greencell}{RGB}{146,210,14}
\definecolor{redcell}{RGB}{250,70,11}

\usepackage{multirow}
\usepackage{comment}
\usepackage{tikz}
\usetikzlibrary{patterns}
\usepackage{xurl}

\definecolor{kvunreused}{HTML}{FFF7E7}

\begin{document}
\title{\sysname: Privacy-Aware and Fine-Grained KV Cache Sharing for Efficient LLM Inference}

\author{Guanlong Wu}
\authornote{Equal contribution}
\email{santiscowgl@gmail.com}
\affiliation{%
  \institution{SUSTech}
}

\author{Zhaohan Li}
\authornotemark[1]
\email{ray.eldath@outlook.com}
\affiliation{%
  \institution{SUSTech}
}

\author{Yao Zhang}
\email{zhangyao.crypto@bytedance.com}
\affiliation{%
  \institution{ByteDance}
}

\author{Zheng Zhang}
\email{aca2hang2heng@gmail.com}
\affiliation{%
  \institution{SUSTech}
}

\author{Jianyu Niu}
\email{Njianyu@gmail.com}
\affiliation{%
  \institution{SUSTech}
}

\author{Ye Wu}
\email{wuye.2020@bytedance.com}
\affiliation{%
  \institution{ByteDance}
}

\author{Yinqian Zhang}
\authornote{Corresponding author}
\email{yinqianz@acm.org}
\affiliation{%
  \institution{SUSTech}
}

\begin{abstract}

Large Language Models (LLMs) rely on Key-Value (KV) caching to accelerate inference, and many serving systems further share the KV cache across users' requests to reduce redundant computation. While widely adopted, unrestricted cross-user sharing introduces side-channel vulnerabilities, allowing an adversary to infer user inputs by probing for cache reuse.
Existing defenses disable sharing entirely to prevent leakage; yet such a coarse-grained strategy sacrifices substantial reuse potential, since prompts often include large portions of privacy-irrelevant segments, such as system instructions or publicly accessible materials.
Building on this, we present \sysname, a privacy-aware KV cache sharing mechanism that enables fine-grained reuse of KV entries across requests.
Realizing such fine granularity requires token-level cache management, as reusable segments vary in length and position due to sensitivity masking, making reuse more complex than the fixed-size or sentence-level chunking used in existing coarse-grained schemes.
Specifically, \sysname makes fine-grained reuse practical by addressing two key challenges: accurately and efficiently deriving reusable KV segments and efficiently retrieving them over variable-length spans.
We implement \sysname on top of vLLM and evaluate it on three datasets, showing that it eliminates direct leakage through KV cache reuse side channels while reducing TTFT by 4.5x and increasing cache hit rates by 44\% compared with state-of-the-art approaches.

\end{abstract}

\settopmatter{printacmref=false}
\makeatletter
\renewcommand\footnotetextcopyrightpermission[1]{}
\makeatother

\maketitle

\section{Introduction}
\label{sec:introduction}

Large Language Models (LLMs)~\cite{touvron2023llama,achiam2023gpt} have achieved remarkable progress in production, with enormous usage across diverse domains and user bases. 
To meet the massive computational demands, numerous optimizations have been developed to enhance inference efficiency, among which one widely-adopted method is to share the Key-Value (KV) cache~\cite{kwon2023efficient,zheng2024sglang} across different users. 
KV cache sharing involves storing and reusing the intermediate attention states computed during inference, allowing multiple requests to leverage previously computed key-value pairs when processing overlapping input sequences. 
This mechanism reduces memory usage and computational overhead, enabling higher system throughput and lower response latency.
KV cache sharing has been extensively deployed in both open-source implementations (\eg, vLLM~\cite{kwon2023efficient}, SGLang~\cite{zheng2024sglang}) and production systems (\eg, OpenAI~\cite{openaipromptcaching}, Claude~\cite{claudepromptcaching}, Gemini~\cite{geminipromptcaching}), demonstrating its practicality in LLM serving infrastructure.

Unfortunately, recent studies~\cite{wu2025know,song2024early,zheng2024inputsnatch} show that unrestricted sharing across mutually untrusted users exposes a new side-channel surface. 
At a high level, an adversary sends probe requests and checks for KV cache reuse through side channel signals such as reduced latency~\cite{song2024early,zheng2024inputsnatch} or scheduling priority~\cite{wu2025know}. 
A reuse signal indicates that the probe matches part of the victim’s prompt, enabling the adversary to reconstruct the victim's input.
Existing defenses (\eg, discussed by NVIDIA~\cite{nvidiakv}) suggest disabling cross-user sharing entirely, which eliminates leakage but also sacrifices most of the efficiency benefits of KV cache sharing.

However, the existing all-or-nothing sharing policy is overly coarse-grained and sacrifices significant reuse potential, as user prompts often contain substantial privacy-irrelevant segments in practice that could be safely shared.
As motivating examples, we first note that prompts often include substantial system and public content, such as instructions, templates, or retrieved passages in retrieval-augmented generation (RAG) pipelines, unrelated to user input; our analysis on the MSMARCO~\cite{msmarco} dataset (\secref{subsec:observations}) shows that user-provided text constitutes only 8.8\% of the total prompt. 
Besides, even within the user-provided portion, most content remains privacy-irrelevant; prior studies~\cite{narayanan2010myths,mccallister2010guide,schwartz2011pii} report that in domains such as banking or customer-service, only personally identifiable information (PII) like names or account numbers is considered sensitive.
Using the SOTA PII identifier Presidio~\cite{presidio}, we found such sensitive tokens account for merely 2.3\% of the user input (\secref{subsec:observations}).

\textbf{Together, these highlight the need for a practical mechanism to enable fine-grained selective KV cache sharing, thereby breaking the all-or-nothing trade-off between privacy and efficiency.}
Nonetheless, existing systems (\eg, vLLM, CacheBlend, CacheCraft, Epic)~\cite{kwon2023efficient,yao2025cacheblend,agarwal2025cache,hu2024epic} operate at coarse granularity and thus fundamentally cannot support selective sharing: they reuse KV cache at the level of fixed chunks (e.g., 512 tokens~\cite{agarwal2025cache}) or entire prompt, so the presence of a single sensitive token (e.g., PII) invalidates the whole unit and discards most otherwise reusable content (as detailed in \secref{subsec:selectivesharing}).

To this end, we propose \sysname, which realizes privacy-aware selective KV cache sharing. 
Specifically, for each prompt after inference, \sysname integrates privacy detection (\eg, user-defined or existing detection tools) to identify and exclude sensitive segments of the prompt, then derives the reusable KV segments from the remaining tokens, storing them for  retrieval across requests.
We identify two key challenges in realizing such fine-grained KV cache sharing beyond existing chunk-level designs:

\begin{packeditemize}

\item \textbf{Challenge 1: Deriving reusable KV segments.}
Achieving fine-grained, privacy-aware sharing goes beyond identifying non-sensitive text, as KV segments are context-dependent and naively reusing them under mismatched contexts leads to quality degradation.
In practice, such reusable segments can also emerge at arbitrary positions and with variable lengths, making their identification and validation a major challenge.
While prior work~\cite{agarwal2025cache} defines reusability using intra- and inter-attention comparisons, simply applying this criterion to fine-grained sharing would require evaluating every possible token pair to compute cross-segment dependencies, leading to an $O(n^2)$ complexity that makes it impractical for real-time inference (\secref{subsubsec:c1}).
To make fine-grained sharing practically viable, we design an adaptive algorithm based on summed-area tables that efficiently evaluates segment reusability without incurring additional computational cost.

\item \textbf{Challenge 2: Efficiently retrieving reusable segments.}
Once reusable KV segments are derived, the system must efficiently match them against incoming requests to enable reuse. 
Existing methods rely on fixed-size segments with predefined boundaries, allowing direct hash-based lookup. 
However, under fine-grained sharing, segments have variable lengths and positions, where even a single-token shift breaks hash alignment and reduces matching to substring search over all spans, leading to prohibitive overhead. 
To address this, we design a rolling-hash–based retrieval algorithm that supports constant-time updates and enables linear-time matching for variable-length segments.

\end{packeditemize}

We implement a prototype of \sysname on top of vLLM~\cite{kwon2023efficient} and conduct comprehensive evaluations. 
Our results demonstrate the effectiveness of \sysname from three perspectives:
\begin{packeditemize}

    \item \textbf{Privacy:} (i) \emph{Within our threat model,} we show that tokens marked as sensitive cannot be directly recovered through reuse-based side channels, eliminating the leakage channel targeted in prior attacks~\cite{wu2025know,song2024early,zheng2024inputsnatch}.  
    (ii) \emph{Beyond our threat model,} we note that an attacker may infer sensitive information from non-sensitive tokens, \ie, contextual leakage, a broader open problem~\cite{yan2024protecting} but not considered in prior attacks~\cite{wu2025know,song2024early,zheng2024inputsnatch}.
    We provide a dedicated evaluation (\secref{subsec:rq0}), showing that contextual leakage remains limited in practice, as \sysname does not expose the positions or counts of sensitive tokens, nor the grouping or ordering of non-sensitive tokens, resulting in only 2.2\% exact recovery and 6.4\% semantic recovery under SOTA attacks~\cite{staab2023beyond}.
    (iii) \emph{Privacy under imperfect privacy detection.} \sysname is decoupled from the underlying detector and is compatible with a wide range of mechanisms (from user-defined rules to detectors), reflecting a common security-by-design practice built on well-established privacy detection (\eg, Presidio~\cite{presidio} or OpenAI Privacy Filter~\cite{openaiprivacyfilter}). However, detectors are inherently imperfect—a limitation shared by all detection-based systems. 
    We analyze detection errors and show that false negatives (unmarked sensitive tokens) increase direct leakage, while false positives (over-marked non-sensitive tokens) reduce contextual leakage (\secref{subsec:rq0}).
    \item \textbf{Generation quality:} \sysname preserves model output quality across datasets and models, showing negligible deviation from full recomputation and aligning with prior works~\cite{yao2025cacheblend,agarwal2025cache,hu2024epic}.
    \item \textbf{Efficiency:} (i) End-to-end evaluations show that \sysname reduces TTFT by up to 4.5x and reuses up to 94.57\% of tokens’ KV cache compared to non-sharing baselines. 
    (ii) Even without privacy constraints (\ie, privacy filtering is disabled), CachePrune’s fine-grained sharing improves cache hit rate by 44\% over existing reuse mechanisms,
    showing the efficiency benefit of fine-grained KV reuse itself. (iii) We further characterize the tradeoff between privacy protection and efficiency: stronger privacy requirements mark more tokens as sensitive, fragmenting prompts into smaller pieces. This fragmentation reduces the likelihood of forming reusable segments, thereby lowering reuse opportunities. 
\end{packeditemize}

\bheading{Our contributions.} To summarize, we make the following contributions in this paper:
\begin{packeditemize}

\item We present \sysname, the first practical KV cache sharing system that breaks the privacy-efficiency trade-off by enabling fine-grained reuse of non-sensitive KV entries, mitigating direct recovery via reuse-based side channels~\cite{wu2025know,song2024early,zheng2024inputsnatch} by restricting reuse of tokens identified as sensitive.

\item We introduce fine-grained KV cache sharing that overcomes the limitations of fixed chunk-based reuse. We show that even without privacy constraints, fine-grained sharing alone improves reuse efficiency over existing approaches.

\item We implement \sysname upon vLLM and conduct comprehensive evaluations. \sysname is compatible with existing LLM serving frameworks (e.g., vLLM) and provides a unified interface for integrating diverse privacy detection mechanisms. Our results demonstrate (i) elimination of direct leakage from reuse-based side channels, (ii) preserved generation quality across tasks and models, and (iii) significant efficiency gains together with a characterization of the privacy–efficiency tradeoff.

\end{packeditemize}

\section{KV Cache}
\label{sec:background}

\subsection{Fundamentals}
\label{subsec:fundamentals}
Large language models (LLMs)~\cite{naveed2023comprehensive,zhao2023survey} are built on the Transformer~\cite{vaswani2017attention}, where each token is represented by a \textbf{K}ey vector (for attention matching) and a \textbf{V}alue vector (for information retrieval).
In autoregressive generation, tokens are processed sequentially, and each step attends to all previous tokens via their \textbf{K}ey and \textbf{V}alue vectors.
To avoid recomputation, modern systems maintain a \textbf{K}ey-\textbf{V}alue (KV) cache~\cite{li2024survey} that stores these vectors across layers, making attention cost grow only with newly generated tokens rather than the full sequence length.
While KV caching is a standard optimization for LLM inference, it shifts the bottleneck from computation to memory~\cite{jin2024compute}.
Caching attention states reduces decoding latency, but storing key and value tensors for all past tokens causes memory usage to scale with sequence length and model size.
In multi-tenant serving, this trade-off directly constrains concurrency and throughput under a fixed GPU budget.

\subsection{KV Cache Sharing}
\label{subsec:kvcachesharing}

\begin{figure}[t]
  \centering

  \begin{minipage}{\linewidth}
    \centering
    \includegraphics[width=\linewidth]{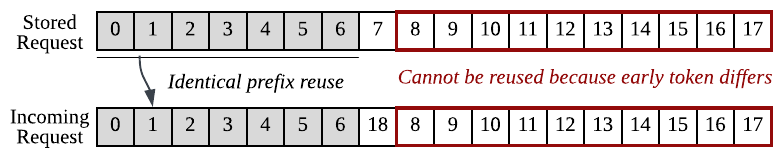}
    {\scriptsize (a) Prefix sharing.}
    \label{subfig:prefixsharing}
  \end{minipage}

  \vspace{0.6em}

  \begin{minipage}{\linewidth}
    \centering
    \includegraphics[width=\linewidth]{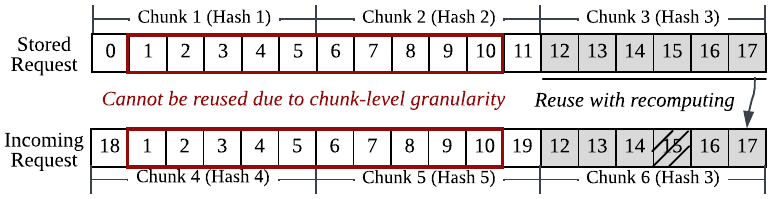}
    {\scriptsize (b) Position-independent sharing.}
    \label{subfig:positionindependent}
  \end{minipage}

  \caption{KV cache sharing mechanisms. White blocks denote unreusable KV cache, gray blocks denote reusable KV cache, and hatched blocks denote recomputed KV cache.}
  \label{fig:kvsharing}
\end{figure}

While KV caching improves efficiency, its memory usage still scales with sequence length and model size, limiting scalability in practice.
KV cache sharing~\cite{kwon2023efficient,zheng2024sglang} addresses this by reusing cached KV across requests with overlapping content, avoiding redundant storage and computation.
This reduces memory footprint, enabling higher concurrency, while also lowering per-request latency.
Existing approaches mainly fall into two categories: prefix sharing~\cite{kwon2023efficient,zheng2024sglang} and position-independent sharing~\cite{yao2025cacheblend,agarwal2025cache,hu2024epic} (\figref{fig:kvsharing}).

\subsubsection{Prefix Sharing}
Prefix sharing reuses cached KV pairs when requests share the same initial sequence.
The system matches the prefix of an incoming request against existing cache entries and reuses the corresponding KV vectors, computing only the remaining tokens.
As in \figref{fig:kvsharing}-a, tokens 0–6 overlap between the stored and incoming requests, enabling KV reuse.
This is the most widely adopted form of KV cache sharing, used in systems such as vLLM~\cite{kwon2023efficient}, SGLang~\cite{zheng2024sglang}, and DeepSeek~\cite{liu2024deepseek}, and deployed in production including OpenAI~\cite{openaipromptcaching}, Anthropic~\cite{claudepromptcaching}, and Gemini~\cite{geminipromptcaching}.

\bheading{Advantages.}
Prefix sharing does not affect model accuracy, as reused KV entries are identical to recomputed ones.

\bheading{Limitations.}
Reuse is limited to exact prefix matches; any difference in early tokens prevents sharing, leaving overlapping sequences unused (\eg, token 8-17 in \figref{fig:kvsharing}-a).

\subsubsection{Position-independent Sharing}

Position independent sharing is an emerging approach designed to increase the usability of KV cache sharing beyond exact-prefix matches. 
Recent work~\cite{yao2025cacheblend,agarwal2025cache,hu2024epic} shows that KV entries can be reused even when overlapping content appears at different positions, with only a slight accuracy trade-off. 
The approach leverages the sparsity of attention in LLMs: only a subset of ``important'' tokens strongly influences each step.
By recomputing the important tokens, many other KV entries can still be reused without degrading output quality.

\figref{fig:kvsharing}-b illustrates the process. 
When a new request arrives, the input is divided into fixed-size chunks (e.g., 512 tokens in practice and 6 tokens in this case). 
Each chunk is matched against the cache using hash-based lookup; matched chunks (e.g., chunk 3 and chunk 6) reuse their KV entries, while a small subset of tokens within them (e.g., token 15) is recomputed to preserve generation quality. 
Representative systems include CacheBlend~\cite{yao2025cacheblend}, CacheCraft~\cite{agarwal2025cache}, and EPIC~\cite{hu2024epic}, which mainly differ in their strategies for selecting tokens to recompute.

\begin{packeditemize}
\item CacheBlend~\cite{yao2025cacheblend} selects tokens for recomputation based on KV deviation. It compares the first-layer KV of a chunk in its original context with that from a full recomputation in the new context, then recomputes the top 15\% of tokens with the largest deviation, as these tokens are most likely to change in attention distribution under the new context.
\item EPIC~\cite{hu2024epic} extends CacheBlend by observing that boundary tokens are most critical for maintaining cross-chunk dependencies. Instead of recomputing a fixed 15\% of high-deviation tokens, it recomputes only the first and last K/2 tokens of the chunk, where K is a predefined value, making the recomputation cost constant.
\item CacheCraft~\cite{agarwal2025cache} provides an explainable methodology for selective recomputation, moving beyond the heuristic nature of KV deviation in EPIC and CacheBlend. It selects tokens directly from attention scores: within a reused chunk, tokens whose attention is concentrated inside the chunk are retained, while those attending more to tokens outside the chunk have their KV recomputed in the new context. This principled approach forms the basis of our system and will be detailed in \secref{subsubsec:c1}. 

\end{packeditemize}

\bheading{Advantages.} Position-independent sharing greatly increases the opportunities for KV reuse compared to prefix sharing.

\bheading{Limitations.}
All existing position-independent methods reuse at fixed chunk granularity: both cached and incoming requests are split into equal-sized chunks and matched by hash lookup. 
This rigid segmentation can split identical segments across different boundaries, making them unreusable.
For instance, in \figref{fig:kvsharing}-b, tokens 1–10 cannot be reused because they span in different chunks.

\subsection{Privacy Concerns in KV Cache Sharing}
\label{subsec:privacyconcern}

While KV cache sharing improves inference efficiency, it also opens up new privacy risks. Because the KV cache encodes intermediate states directly derived from user prompts, unrestricted sharing across mutually distrustful users creates a side channel that can be exploited to leak sensitive input.

\bheading{Attack procedure.}
Existing attacks~\cite{zheng2024inputsnatch,wu2025know,song2024early} follow the same high-level procedure, where the attacker interacts with the LLM service as a normal client:
To test whether a victim’s prompt contains a certain token sequence, the attacker issues a probe request containing that sequence. 
If the KV cache already holds matching entries from the victim’s request, the attacker’s probe benefits from cache reuse. 
This reuse can be detected through observable side effects such as reduced response latency~\cite{song2024early,zheng2024inputsnatch} or higher scheduling priority in cache-aware systems~\cite{wu2025know}. 
By repeating this process with different probes and tracking which ones trigger reuse, the attacker can gradually reconstruct the victim’s prompt.

\bheading{Existing defense.} A straightforward defense is to block KV cache sharing or restrict it within the same user session only~\cite{nvidiakv,pang2024cache}. 
However, this drastically reduces cache hit rates, increases memory usage, and nullifies much of the latency and throughput improvements that make KV caching attractive in the first place.
This motivates our work, which aims to preserve the performance benefits of KV cache sharing while ensuring its security, as detailed in \secref{sec:motivation}.

\section{Motivation}
\label{sec:motivation}

\subsection{Problem Statement}
\label{subsec:problemstatement}
KV cache sharing is widely used in multi-tenant LLM serving to reduce redundant computation (\secref{subsec:kvcachesharing}).
However, since cached states are derived from user prompts, cross-user sharing exposes side-channel signals that allow adversaries to infer other users’ inputs (\secref{subsec:privacyconcern}). 
Existing defenses~\cite{nvidiakv} eliminate this by disabling cross-user sharing, as such attacks fundamentally rely on detecting reuse. 
While effective, this removes most of the efficiency gains of KV caching: as shown in \figref{fig:ttftreduction}, TTFT can increase by up to 14x for a 10,000-token input on Mistral-7B~\cite{mistral}.
Given that prompts often contain substantial non-private content (\secref{subsec:observations}), we aim to enable cross-user sharing while mitigating side-channel leakage (\secref{subsec:selectivesharing}).

\bheading{Threat model.} We summarize our threat model, following prior work~\cite{wu2025know,song2024early,zheng2024inputsnatch}, as follows:
\begin{packeditemize}
    \item \textit{Adversary's capabilities.} The adversary is a client-side user with the same privileges as any ordinary user: it can only interact with the service via standard client APIs, with no direct access to the server, model weights, or plaintext KV cache entries. 
    Its background knowledge is limited to the publicly available tokenizer of LLM services. 
    We model the adversary’s observation as a binary reuse oracle $R(p) \in \{0,1\}$, where $R(p) = 1$ indicates that a probe prompt $p$ triggers reuse of cached KV entries, and $R(p) = 0$ otherwise; such oracles can be inferred indirectly through side channels such as latency reduction or scheduling priority. 
    By issuing a sequence of probe requests $P = \{ p_1, p_2, \dots, p_k \}$ and observing the corresponding reuse oracles $R(p_i)$, the adversary can infer the presence of specific token sequences in other users’ prompts.

    \item \textit{Adversary's goals.} 
    Identical to prior work~\cite{wu2025know,song2024early,zheng2024inputsnatch}: the adversary aims to directly recover private content in other users’ prompts \textbf{solely} through side channels from KV cache sharing. 
    Formally, let a prompt be a token sequence \( P = \{ t_1, t_2, \dots, t_n \} \), and let \( M \in \{0,1\}^n \) be a binary sensitivity mask, where \( M_i = 1 \) indicates that token \( t_i \) is sensitive. The adversary’s goal is to recover tokens or spans with \( M_i = 1 \) based only on observed reuse oracles.

\end{packeditemize}

\bheading{Design goals.} 
We aim to build a practical KV cache sharing system that breaks the privacy–efficiency trade-off by enabling fine-grained reuse of non-sensitive KV entries, and preventing direct recovery of tokens identified as sensitive through reuse-based side channels.

\iheading{\emph{Out of scope.}} Sensitive information may be inferred from non-sensitive tokens, \ie, contextual leakage, a broader open problem inherent to fine-grained text processing~\cite{yan2024protecting,staab2023beyond}. 
In addition, privacy identification relies on automated detectors~\cite{presidio,openaiprivacyfilter} that may be imperfect, a limitation shared by all detector-based systems. 
While treated orthogonal to our design, we provide dedicated evaluations (\secref{subsec:rq0}) to quantify their residual privacy impact in our setting.

\subsection{Observations}
\label{subsec:observations}

\begin{figure}[t]
    \centering
    \begin{minipage}{0.5\linewidth}
        \centering
        \includegraphics[width=\linewidth]{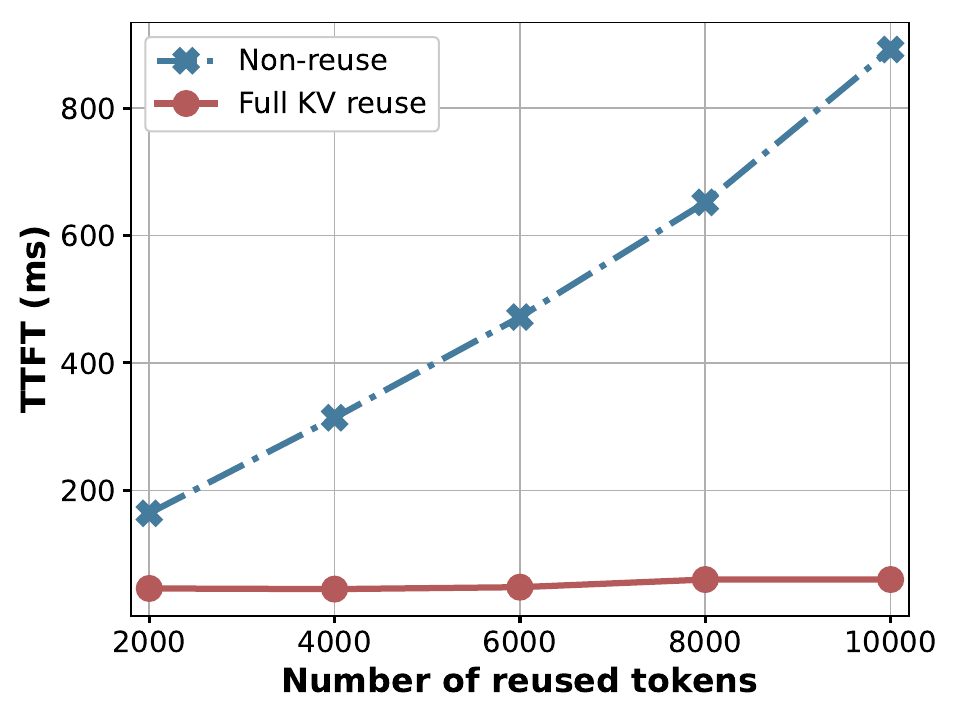}
        \caption{TTFT reduction.}
        \label{fig:ttftreduction}
    \end{minipage}
    \hfill
    \begin{minipage}{0.46\linewidth}
        \centering
        \includegraphics[width=\linewidth]{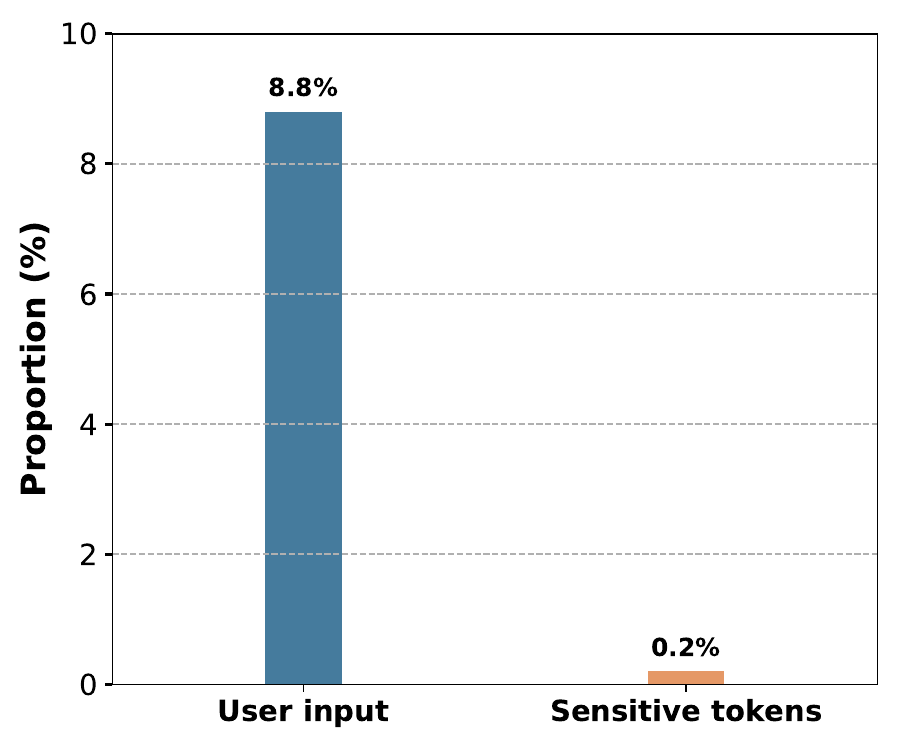}
        \caption{Non-sensitive text.}
        \label{fig:non-sensitive}
    \end{minipage}
\end{figure}

\bheading{Prompts contain abundant privacy-irrelevant content.}
We identify two representative scenarios to illustrate:
\begin{packeditemize}

\item \textbf{System and public context dominates most prompts.}
Prompts are largely composed of system instructions, templates, or retrieved passages (\eg, RAG), while user-provided text forms only a small fraction.
As in \figref{fig:non-sensitive}, analysis on MSMARCO~\cite{msmarco} shows that user-provided content accounts for only 8.8\% of the total prompt.

\item \textbf{Even within user input, most content is inherently privacy-irrelevant.}
In many applications (\eg, financial)~\cite{narayanan2010myths,mccallister2010guide, schwartz2011pii}, only personally identifiable information (PII) of user text is privacy-relevant, while the rest is non-sensitive.
Other works~\cite{nissenbaum2018respecting, ackerman2001privacy} have discussed cases involving context privacy, where the sensitivity of context is largely domain-dependent; for example, under clinical settings, questions about a patient’s symptoms are treated as private, whereas unrelated queries (e.g., weather conditions) in the same setting do not constitute privacy.
Using the Presidio~\cite{presidio} detector, we find that sensitive tokens account for 2.3\% of user input and only 0.2\% of the full prompt (\figref{fig:non-sensitive}).

\end{packeditemize}

\bheading{Opportunity: selectively sharing privacy-irrelevant KV entries.}
These observations point to a middle ground between full sharing and complete isolation.
By restricting cross-user KV cache sharing to tokens that are privacy-irrelevant, the side-channel signal from sensitive tokens is eliminated, while the majority of KV entries remain reusable. 
This preserves privacy guarantees without incurring the large performance penalties of existing defense.

\subsection{Selective KV Cache Sharing}
\label{subsec:selectivesharing}

\bheading{Definition.}
Selective KV cache sharing enables cross-user reuse only for tokens marked as non-sensitive under the sensitivity mask defined in \secref{subsec:problemstatement}. 
Specifically, KV entries for token $t_i$ are reused across users if and only if $M_i = 0$, while tokens with $M_i = 1$ are excluded from cross-user sharing.

\bheading{Privacy protection.}
Selective KV cache sharing guarantees that tokens marked as sensitive cannot be directly leaked through side channels. 
Side-channel attacks fundamentally rely on detecting whether a token’s KV entries are reused across requests (\secref{subsec:privacyconcern}). 
By excluding sensitive tokens from cross-user reuse, their KV entries are never shared and thus produce no reuse signal, rendering direct extraction by side-channel attacks ineffective (aligning with the threat model in \secref{subsec:problemstatement}). 
This applies only to direct leakage; contextual leakage from partially recovered content lies outside this design, yet we evaluate and analyze its impact in \secref{subsec:rq0}.

\bheading{Practical use cases.}
Selective KV cache sharing can flexibly support different levels of privacy protection, depending on how private contents are identified. 
For example, aligning with \secref{subsec:observations}, in a low privacy demand setting, only specific tokens containing personally identifiable information are excluded from reuse, while in a high privacy demand setting, the entire user input is masked and only system prompts or public materials are shared.

\subsubsection{Gaps in Existing Work for Selective KV Cache Sharing}
\label{subsec:gaps}

Selective KV cache sharing cannot be realized by simply extending prior work (\ie, prefix or positional-independent sharing, as depicted in \secref{subsec:kvcachesharing}), since both lead to excessive loss of reusable KV cache under selective-sharing constraints.

\begin{figure}[t]
  \centering

  \begin{minipage}{\linewidth}
    \centering
    \includegraphics[width=\linewidth]{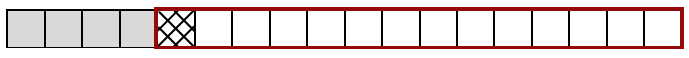}
    {\scriptsize (a) Prefix sharing with selective reuse.}
    \label{subfig:prefixsharinggaps}
  \end{minipage}

  \vspace{0.6em}

  \begin{minipage}{\linewidth}
    \centering
    \includegraphics[width=\linewidth]{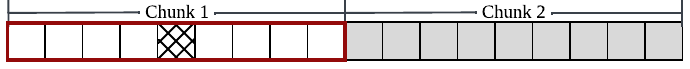}
    {\scriptsize (b) Position-independent sharing with selective reuse.}
    \label{subfig:positionindependentgaps}
  \end{minipage}

    \vspace{0.6em}

  \begin{minipage}{\linewidth}
    \centering
    \includegraphics[width=\linewidth]{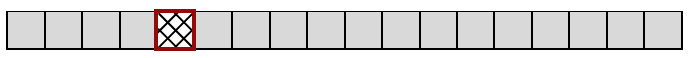}
    {\scriptsize (c) Selective KV cache sharing at token granularity.}
    \label{subfig:tokenlevelselective}
  \end{minipage}

  \caption{Selective KV cache sharing under different mechanisms. White blocks denote unreusable KV cache, gray blocks denote reusable KV cache, and hatched blocks denote sensitive tokens' KV cache.}
  \label{fig:kvsharingOne}
\end{figure}

\bheading{Prefix sharing.}
Prefix sharing requires two sequences to match from the beginning to enable reuse. 
Under selective sharing, once a sensitive token appears, all subsequent tokens become non-reusable. 
Otherwise, reusing any later token would imply that the entire prefix, including the sensitive token, matches across users, thereby leaking its presence (as shown in \figref{fig:kvsharingOne}-a).

\bheading{Position-independent sharing.}
Position-independent sharing also breaks down under selective constraints because it reuses KV entries at the fix-sized chunk level. 
If a single sensitive token appears in a chunk, the entire chunk must be excluded from reuse; otherwise, it would reveal the presence of the sensitive token (as depicted in \figref{fig:kvsharingOne}-b).
Since chunks are typically large (e.g., 512 tokens or more in practice~\cite{agarwal2025cache}), this coarse granularity causes many non-sensitive tokens to be discarded along with the sensitive one, leading to a substantial loss of reuse.

\subsubsection{Need for KV Cache Sharing at Token Granularity}

The shortcomings of prior work show that to fully preserve reuse under selective sharing, KV cache management needs to operate at token granularity. 
In this setting, reuse decisions are made for each token individually, rather than in fixed-size chunks. 
When a token is marked as sensitive, its KV entries are excluded from reuse, while the surrounding tokens remain unaffected. 
Sensitive tokens therefore serve as natural boundaries that partition the prompt into dynamic-length reusable segments. 
These segments can be reused independently without leaking sensitive information  (\figref{fig:kvsharingOne}-c).

Notably, token-level KV cache management not only enables selective sharing for privacy but also improves efficiency in general KV cache management.
In chunk-level management, reusable segments that span across chunk boundaries often cause mismatches and are discarded, reducing cache hit rates (\figref{fig:kvsharing}-b). 
Token-level management eliminates this limitation by deciding reuse at the token level, allowing all matching tokens to be reused regardless of alignment and thereby maximizing cache utilization (\secref{subsec:endtoend}).

\section{System Design of \sysname}
\label{sec:system}

\subsection{Architecture}

\begin{figure}[t]
    \centering
    \includegraphics[width=0.49\textwidth]{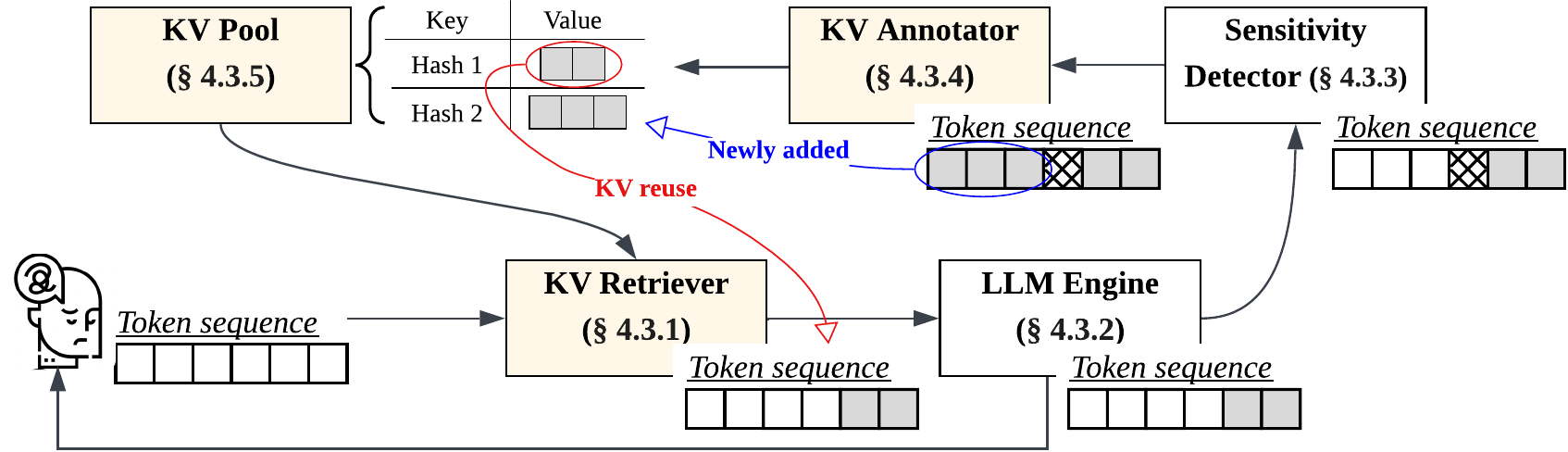}
    \caption{System architecture of \sysname. New modules introduced by \sysname are highlighted in light yellow.}
    \label{fig:architecture}
\end{figure}

\bheading{Components.}
\sysname (\figref{fig:architecture}) consists of five components:

\begin{packeditemize}
\item \textbf{KV Retriever (\secref{subsec:kvretriever}).} 
The \texttt{KV Retriever} searches the \texttt{KV Pool} for KV segments that match the incoming request’s context and links them to the sequence, leaving unmatched positions for the \texttt{LLM Engine} to compute.

\item \textbf{LLM Engine (\secref{subsec:llmengine}).}
The \texttt{LLM Engine} completes missing KV entries, recomputes those marked by the \texttt{KV Annotator}, and generates the final output from the full KV cache. 
We do not modify the functionality of existing LLM engines, and \sysname is compatible with existing frameworks such as vLLM~\cite{kwon2023efficient}.

\item \textbf{Sensitivity Detector (\secref{subsec:sendetector}).} 
The \texttt{Sensitivity Detector} acts as a pluggable interface that identifies privacy-related content within user prompts and marks the corresponding regions for exclusion from KV sharing. 
Such privacy information can be specified through various means, including manual annotations, rule-based matching, or existing detection models. 
While we do not introduce new detection methods, we integrate these approaches to support different levels of privacy guarantees.

\item \textbf{KV Annotator (\secref{subsec:kvannotator}).} 
The \texttt{KV Annotator} identifies reusable KV segments after excluding sensitive KV, and marks KV entries that should be recomputed to maintain LLM generation quality.

\item \textbf{KV Pool (\secref{subsec:kvpool}).} 
The \texttt{KV Pool} stores reusable KV cache segments, indexing them via hashing for fast retrieval, evicting entries when the storage capacity is reached, and resolving conflicts when multiple KV segments overlap.

\end{packeditemize}

\bheading{Workflow.} Upon a new request, the \texttt{KV Retriever} queries the \texttt{KV Pool}, integrates matched segments into the sequence, and passes it to the \texttt{LLM Engine}, which computes missing KV entries and returns the response. Afterward, \sysname performs a post-inference phase (without affecting latency): the \texttt{Sensitivity Detector} identifies sensitive tokens and blocks their sharing, and the \texttt{KV Annotator} derives reusable segments from the remaining tokens while marking entries requiring recomputation. The resulting segments and metadata are stored in the \texttt{KV Pool} for future reuse.

\subsection{Challenges and Algorithmic Solutions}
\label{subsec:challenges}

\subsubsection{C1: Efficiently and Accurately Deriving Reusable KV Segments} 
\label{subsubsec:c1}

After inference, the \texttt{Sensitivity Detector} analyzes the input token sequence \( P = \{ t_1, t_2, \dots, t_n \} \) and produces a binary sensitivity mask \( M \in \{0,1\}^n \), where \( M_i = 1 \) indicates that token \( t_i \) contains sensitive information. 
Tokens with \( M_i = 1 \) act as separators, partitioning \(P\) into coarse-grained segments \( S_1, S_2, \dots, S_k \), where each segment 
\( S_m = \{ t_{a_m}, t_{a_m+1}, \dots, t_{b_m} \} \) satisfies \( M_{a_m-1} = 1 \) or \(a_m = 1\), and \( M_{b_m+1} = 1 \) or \(b_m = n\).
However, such segments are not directly reusable, since each token’s KV state depends on all preceding tokens in the original context.
Reusing a segment without adjustment introduces mismatched attention dependencies, degrading generation quality (\secref{subsec:kvcachesharing}).

\begin{figure}[t]
    \centering
    \includegraphics[width=0.48\textwidth]{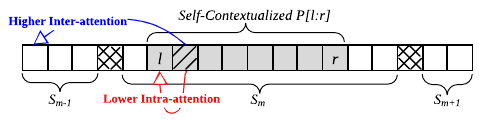}
    \caption{Deriving reusable KV segments under sensitive tokens.}
    \label{fig:selfcontextualize}
\end{figure}

\bheading{Reuse criterion.}
Prior work~\cite{agarwal2025cache} has proposed that a segment is considered reusable if it is \textbf{\textit{self-contextualized}}, \ie, its \textbf{\textit{intra-attention}} (attention paid within the segment) exceeds its \textbf{\textit{inter-attention}} (attention paid to tokens outside the segment). 
Specifically, let \(Attn(i,j)\) denote the attention score from token \(t_i\) to token \(t_j\). 
For a substring \(P[l:r]\) within segment \(S_m\), we define:
\begin{equation*}
\mathrm{IntraAttn}(l,r),\ \mathrm{InterAttn}(l,r)
= \sum_{i=l}^{r} \Big(
\sum_{j=l}^{i} \mathrm{Attn}(i,j),\ 
\sum_{j=1}^{l-1} \mathrm{Attn}(i,j)
\Big)
\end{equation*}
A substring \(P[l:r]\) is therefore reusable if \({IntraAttn}(l,r) > {InterAttn}(l,r)\). 
Moreover, within such a segment, tokens whose \({InterAttn}() - {IntraAttn}()\) is high are marked for recomputation when the segment is reused in a new context.
Thus, \texttt{KV Annotator} aims to refine each coarse segment $S_m$ by selecting the substring $P[l:r] \subseteq S_m$ with the largest intra--inter difference and marking tokens that depend on outside context for 
recomputation (\figref{fig:selfcontextualize}).

However, prior works restrict reuse to fixed chunks with predetermined boundaries (\secref{subsec:kvcachesharing}), whereas token-granularity requires exploring all substrings, dramatically increasing the computational cost.
Specifically, in the chunk-level setting with constant chunk size $c$, the number of candidate segments is $O(n/c)$. 
For each chunk, computing $\mathit{IntraAttn}()$ takes $O(c^2)$ time and $\mathit{InterAttn}()$ takes $O(n)$ (as equation above), 
so the total complexity is $O(n/c \cdot (c^2+n)) = O(n^2)$. 
In contrast, token-granularity allows arbitrary start and end indices $(l,r)$, yielding $O(n^2)$ substrings. 
Evaluating intra- and inter-attention for each still costs $O(n^2)$ from raw scores. 
This naive extension therefore incurs \(O(n^2) \times O(n^2) = O(n^4)\) total time complexity, which is prohibitive in practice.

\begin{figure}[t]
    \centering
    \begin{minipage}{0.23\textwidth}
        \centering
        \includegraphics[width=\linewidth]{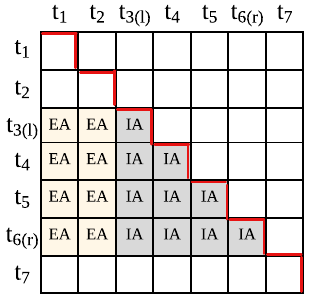}
        {\scriptsize (a) Attention matrix}
    \end{minipage}
    \hfill
    \begin{minipage}{0.23\textwidth}
        \centering
        \includegraphics[width=\linewidth]{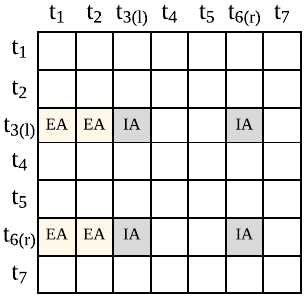}
        {\scriptsize (b) Summed area table}
    \end{minipage}

    \caption{Example of intra- (IA) and inter-attention (EA) computation using the attention matrix and the summed area table, with $t_{3}$ and $t_{6}$ as the boundaries $l,r$.}
    \label{fig:scorecalculate}
\end{figure}

To this end, we propose a method based on the \textit{summed area table}~\cite{crow1984summed,hensley2005fast}, a widely adopted algorithm that supports constant-time retrieval of sub-region sums after an $O(n^2)$ preprocessing step. 
By applying this to the attention score matrix, we obtain intra/inter attention values for any candidate substring in \(O(1)\) time, reducing the overall search from \(O(n^4)\) to \(O(n^2)\). 
The process has three steps:

\bheading{Step 1: Constructing the summed area table in-place.}  
Let $A \in \mathbb{R}^{n \times n}$ denote the attention score matrix, where $A[i,j]$ is the attention weight from token $t_i$ to token $t_j$. 
We overwrite $A$ in place to store its summed area table $T$, where each entry $T[i,j]$ contains the sum of all elements in the submatrix $[1\!:\!i] \times [1\!:\!j]$:
\[
T[i,j] \gets A[i,j] + T[i-1,j] + T[i,j-1] - T[i-1,j-1]
\]
with any term having an index \(< 1\) treated as zero.  
This preprocessing requires one update per request, runs in \(O(n^2)\) time and uses \(O(1)\) additional memory, since \(A\) is updated in place.
Once \(T\) is computed, the sum of any rectangular subregion \(R = [x_1, x_2] \times [y_1, y_2]\) of the original attention matrix can be retrieved in \(O(1)\) time:
\[
\mathrm{sum}(R)=T[x_2,y_2]-T[x_1\!-\!1,y_2]-T[x_2,y_1\!-\!1]+T[x_1\!-\!1,y_1\!-\!1]
\]
Here, out-of-bound indices are treated as zero. 
Since intra- and inter-attention are sums over contiguous rectangular regions in $A$, this formula yields them in $O(1)$ (\figref{fig:scorecalculate}).

\bheading{Step 2: Optimal substring search within coarse segments.}  
With the summed area table in place, we can now evaluate intra-inter attention for any substring in $O(1)$ time. 
Given a coarse segment $S_m$ from the \texttt{Sensitivity Detector}, we enumerate all substrings $P[l,r] \subseteq S_m$ and compute their intra–inter difference. 
The substring with the maximum difference is selected as the reusable segment. 
Since there are $O(n^2)$ candidates within a segment, the overall search runs in $O(n^2)$ time with no extra space overhead.

\bheading{Step 3: Annotating tokens for recomputation.}
For the selected reusable substring $[l^\ast,r^\ast]$ from step 2, we compute for each token $t_i$ its inter-attention (to tokens before $l^\ast$) and intra-attention (to tokens within $[l^\ast,i]$), both obtained in $O(1)$ time from the summed area table. 
Tokens are then ranked by their inter–intra difference, and the top $\rho\%$ (e.g., 25\% in \secref{subsubsec:recompute}) are marked for recomputation.

\bheading{Optimization.}  
In our experiments, we observe that the length of a reusable segment strongly correlates with the generation quality it can preserve.  
To avoid redundant computation on short substrings with limited contribution, we introduce a configurable minimum segment length parameter (e.g., 128 tokens, as evaluated in \secref{subsubsec:length}) for a segment to qualify as reusable.

\subsubsection{C2: Retrieving Dynamic-Length KV Segments} 
\label{subsubsec:c2}
Existing KV retrieval methods (\secref{subsec:kvcachesharing}) rely on fixed-size chunks, where matching reduces to $O(1)$ hash lookups per chunk. 
In contrast, \sysname operates at token granularity, producing variable-length segments (\secref{subsubsec:c1}). 
Retrieval therefore becomes a substring containment problem: identifying whether a stored segment appears as a contiguous substring of the request. 
This cannot be handled by direct hash lookup; a naïve solution that checks all substrings incurs $O(nm)$ complexity, where $n$ and $m$ are the lengths of the request and segment. 
Since this retrieval step must be performed before generating a response for every request, such overhead would be incurred for all queries, making it too slow for practical deployment.

\begin{figure}[t]
    \centering
    \includegraphics[width=0.48\textwidth]{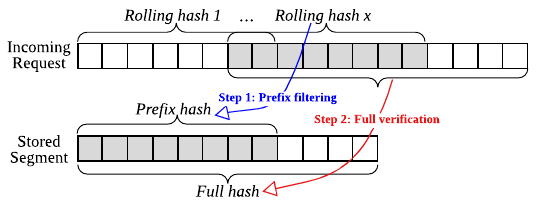}
    \caption{Retrieving KV segments based on rolling hash.}
    \label{fig:rollinghash}
\end{figure}

To this end, we propose an efficient retrieval algorithm for dynamic-length cached segments based on rolling hash. 
Rolling hash~\cite{chayapathi2021survey,jiang2020rolling} can compute the hash of all fixed-size substrings (\ie, window) by calculating the first in full and then updating subsequent ones in $O(1)$ time as the window shifts.
This allows us to scan an incoming request for potential matches to cached segments in near-linear time, a substantial improvement over the naive $O(n \times m)$ approach. 
The algorithm proceeds in two steps (\figref{fig:rollinghash}):

\bheading{Step 1: Prefix filtering.}
For each cached segment we precompute and store the hash of its prefix (by default first 128 tokens, as stated in \secref{subsubsec:c1}, every segment is at least 128 tokens). 
When a new request of length $n$ arrives, we slide a 128-token window across it, updating the hash in $O(1)$ per shift, and compare each window’s hash to the stored 128-token prefix hashes. 
This produces a small set of candidates while scanning the request in $O(n)$ time.

\bheading{Step 2: Full verification.} 
For each candidate, we verify the match over the full segment of length $m$.
A prefix-hash array enables computing the rolling hash of the corresponding substring in $O(1)$ time for a fast pre-check against the cached hash.
To ensure correctness, we then compute a cryptographic hash (\ie, SHA-256) of the substring and compare it with the stored hash, eliminating collision-induced false matches without token-by-token comparison.

\bheading{Cost analysis.}
This design reduces the worst-case $O(n \times m)$ cost of naive substring search to $O(n + c)$, where $c$ is the number of candidates identified by the prefix filter, typically small in practice. 
The space overhead is modest, requiring only the prefix-hash array for the incoming request, precomputed powers, and two hash values per cached segment (prefix and full length).
Our evaluation in \secref{sec:evaluation} shows that the average cost is only 8 ms.

\bheading{Choice of rolling hash algorithm.}
We use polynomial rolling hash modulo the 64-bit Mersenne prime $2^{61}\!-\!1$ with a random base. 
It supports constant-time substring hashing via prefix hashes with precomputed powers, which is critical for efficient verification. 
In contrast, alternatives such as buzhash or rolling CRC support only fixed-window updates and require $O(m)$ time to hash an arbitrary substring, increasing verification cost and reducing performance.

\subsection{Component-Level Design}

\subsubsection{KV Retriever}
\label{subsec:kvretriever}

The \texttt{KV Retriever} fetches KV segments from the \texttt{KV Pool} and integrates them into the request sequence.

\bheading{Integrating multiple sharing policies.}  
The \texttt{KV Retriever} supports both same-user and cross-user sharing.  
When a new request arrives, the retriever first checks the session identity.  
If the request belongs to the same user session, all KV cache can be reused without restriction.  
If the request originates from a different user, it applies the selective cross-user sharing policy, which reuses only non-sensitive segments based on the method described in \secref{subsubsec:c2}.

\bheading{Operational details.}
We separate control logic from data storage: matching and verification are performed on the CPU using lightweight metadata, while full KV tensors remain on the GPU to avoid expensive transfers.  
Metadata consists of token spans, hash values, and indices, which allow the \texttt{KV Retriever} to identify and link reusable segments without touching the actual KV.  
Unmatched or recomputation-marked tokens are represented as zero placeholders, ensuring a consistent sequence structure for the \texttt{LLM Engine}, which later fills them during normal inference.

\subsubsection{LLM Engine}
\label{subsec:llmengine}
The \texttt{LLM Engine} fills missing KV entries and generates the output. 
\sysname does not modify the inference backend and only requires changes to the KV cache management interface, making it compatible with existing frameworks.

\subsubsection{Sensitivity Detector}
\label{subsec:sendetector}

The \texttt{Sensitivity Detector} identifies sensitive content and excludes the corresponding KV entries from reuse. 
\sysname leverages existing mechanisms rather than introducing new ones, supporting:

\bheading{Policy 1: User-defined sensitivity.}
Users can specify sensitive tokens directly through custom rules, \eg, by defining regex patterns for phone numbers or credit card numbers. 
This method gives users explicit control over what information should be protected, and the detector marks all matches in the request as sensitive.

\bheading{Policy 2: Classifier-based detection.}
The detector can also integrate existing classifiers, such as NER models~\cite{mohit2014named} or LLMs~\cite{ponomarenko2026capid}, to automatically identify sensitive entities including names, locations, or organizations. 
This approach is more flexible and accurate, capturing sensitive information under various context.

\bheading{Policy 3: Strict masking policy.}
For applications that require stronger guarantees, we support a stricter mode in which all user-provided input is masked as sensitive by default. 
In this setting, only fixed system prompts or external passages (e.g., retrieved from a knowledge base) are considered safe to share. 
This ensures that no portion of the user’s query is ever reused across sessions, eliminating the risk of leaking private input.

\subsubsection{KV Annotator}
\label{subsec:kvannotator}

The \texttt{KV Annotator} derives reusable segments and marks tokens that require recomputation.

\bheading{Operational details.}
After inference, the \texttt{KV Annotator} transfers the attention score matrix from GPU to CPU memory and applies the algorithm described in \secref{subsubsec:c1} to determine both reusable segments and recomputation masks. 
The calculation is performed on the attention scores of the final transformer layer only, as prior work~\cite{vig2019analyzing,bhojanapalli2021leveraging,ferrando2022measuring} has shown that the last layer sufficiently captures the contextual dependencies relevant for reuse. 

\subsubsection{KV Pool}
\label{subsec:kvpool}

The \texttt{KV Pool} manages the storage and lifecycle of reusable KV segments, defining how they are stored and evicted.

\bheading{Storage format and indexing.}
Each stored KV segment maintains its token sequence along with two hash values: a prefix hash and a full-segment hash. 
These indices are stored in CPU memory, while the actual KV cache values are retained in GPU memory for fast access during inference. 
We acknowledge that some prior works propose offloading KV cache values themselves to CPU memory to reduce GPU usage~\cite{yao2025cacheblend}. 
Such techniques are orthogonal to our design and can be integrated in future extensions.

\bheading{Lifecycle management.}
The pool ensures efficient storage by keeping only one copy of duplicated segments: if one sequence is a strict subsequence of another (e.g., sequence \(A\) contains \(B\)), only the longer segment is preserved. 
When GPU memory reaches capacity, cached segments are evicted using a least recently used (LRU) policy, ensuring that frequently accessed segments are retained.

\section{Evaluation}
\label{sec:evaluation}

We evaluate \sysname, addressing four main research questions:

\begin{packeditemize}
    \item \textbf{[RQ1] Defense effectiveness:} How effectively does \sysname prevent direct leakage via reuse-based side channels (\secref{subsec:problemstatement}), and what level of contextual leakage may still occur? Further, how do these properties change under imperfect privacy detection?

    \item \textbf{[RQ2] Generation quality:} 
    How accurately does \sysname preserve the generation quality, and what are the determining factors that influence the generation quality?

    \item \textbf{[RQ3] Efficiency:}
    How efficient is \sysname compared to non-sharing approaches, and what factors determine its efficiency?

    \item \textbf{[RQ4] Impact of privacy detection policies:}
    How do different privacy detection policies influence accuracy and efficiency, and what trade-offs arise between stronger privacy protection and system performance?
\end{packeditemize}

\subsection{Experimental Setup}
\label{subsec:setup}

\bheading{System configurations.}
We evaluate \sysname across multiple models of varying sizes, including Mistral-7B~\cite{mistral}, Qwen2.5-7B~\cite{qwen7b}, and Qwen2.5-14B~\cite{qwen14b}.
All experiments are conducted on a single NVIDIA H100 GPU with 80 GB of HBM memory. 
The host machine has two Intel Xeon Silver 4510 processors (each 12 cores, 24 threads), 250 GiB of main memory, and a 480 GB SATA SSD supporting a read throughput of 550 MB/s. The CPU and GPU are connected via PCIe 5.0x16, offering a bi-directional bandwidth of 64 GB/s.

\bheading{Datasets.}
We evaluate \sysname on three types of datasets, following the setups used in prior work~\cite{yao2025cacheblend}, and all three datasets provide human-labeled ground truth.
\begin{packeditemize}
    \item \textit{QASPER~\cite{qasper}.} This is a document-grounded question answering dataset based on scientific NLP papers. It includes 5,049 question-answer pairs across 1,585 full-text papers.
    \item \textit{NarrativeQA~\cite{narrativeqa}.} This is a reading comprehension dataset over long narrative texts, \eg movie scripts. It includes 45,987 question–answer pairs over 1,572 stories. 
    \item \textit{QMSum~\cite{qmsum}.} This is a query-based meeting summarization dataset derived from multi-domain meeting transcripts. It contains 232 meetings and 1,808 query–summary pairs.
\end{packeditemize}

\bheading{Evaluation metrics.}
We adopt three categories of metrics:

\begin{packeditemize}

\item \textit{Privacy metrics.}
We evaluate the effectiveness of \sysname in preventing privacy leakage under side-channel attacks using two metrics:
(i) \textbf{Exact recovery rate.}
This measures the fraction of sensitive tokens that can be directly reconstructed by the adversary through KV cache reuse signals, requiring exact token-level match. This aligns with our primary security objective (\secref{subsec:problemstatement}) and prior work~\cite{wu2025know}, where the adversary aims to recover sensitive tokens via side-channel signals.
(ii) \textbf{Contextual recovery rate.}
This measures the extent to which sensitive content can be inferred from recovered non-sensitive tokens (\ie, contextual leakage), which goes beyond our primary threat model. We consider two forms of recovery: (a) exact reconstruction, where sensitive tokens are fully recovered from context, and (b) semantic reconstruction, where the inferred content is not identical but semantically equivalent, as determined by an LLM-based similarity judge~\cite{staab2023beyond}. 

\item \textit{Quality metrics.} 
We adopt two metrics following previous work~\cite{yao2025cacheblend}. 
(i) \textbf{F1-score} measures the word-level overlap between the model output and the ground truth and is suitable for question answering tasks; we use it on QASPER and NarrativeQA. 
(ii) \textbf{ROUGE-L} measures sequence-level similarity via the longest common subsequence and suitable for summarization tasks; we use it on QMSum.

\item \textit{Efficiency metrics.} We report Time-to-First-Token (TTFT) and system throughput as the main indicators of efficiency. 
In addition, we provide a detailed breakdown of the costs, in terms of time, contributed by each system component in \sysname.
\end{packeditemize}

\bheading{Baseline.}
Our primary goal is to break the privacy-efficiency trade-off in KV cache sharing. 
We adopt two types of baselines.
\begin{packeditemize}

\item \textbf{Secure settings.} For the primary evaluation (\secref{subsec:rq0} to \secref{subsec:endtoend}), we compare against a \textit{no-sharing} baseline, where each request is processed independently and the KV cache is not shared across requests. 
This baseline represents the existing secure deployment, as other KV sharing mechanisms do not provide privacy guarantees. 
It worth noting that intra-tenant sharing requires accurate user-level data partitioning and distribution, which is not available in public datasets, making it unable to evaluate.

\item \textbf{Non-secure settings.} To evaluate the efficiency benefit of fine-grained reuse itself without privacy constraints, we compare against representative coarse-grained KV sharing methods, including EPIC, CacheCraft, and CacheBlend (\secref{subsec:minorcontri}). 
These methods enable cross-request reuse but operate at fixed chunk granularity.
\end{packeditemize}

\subsection{RQ1: Defense Effectiveness}
\label{subsec:rq0}

We evaluate \sysname under the threat model in \secref{subsec:problemstatement}, where the adversary aims to recover tokens marked as sensitive solely via reuse-based side channels, and examine whether the attack can be mitigated. We also assess contextual leakage, where sensitive content is inferred from non-sensitive tokens, and the impact of imperfect privacy detection, to quantify residual privacy risks.

\bheading{Methodology.}
We construct the evaluation set by randomly sampling 200 instances from each of the three datasets. For each sample, we use a SOTA privacy detection tool (Presidio~\cite{presidio}) to identify sensitive tokens for \sysname.
We follow a representative side-channel probing attack (\eg, PromptPeek~\cite{wu2025know}). As discussed in \secref{sec:background}, prior KV-cache side-channel attacks share the same inference mechanism and differ only in the observable signal; thus evaluating one representative instantiation suffices to capture this attack class.
We assume a strong adversary with unlimited probing capability. This assumption is motivated by two observations from prior work: (i) adversaries can leverage learning-based strategies (e.g., reinforcement learning) to efficiently explore the input space~\cite{wang2026optileak}, and (ii) even partial background knowledge about user inputs can significantly reduce the probing complexity~\cite{wu2025know,song2024early}. 
Therefore, we do not restrict the adversary’s resources, and treat the results as an upper bound on attack capability. 

Beyond exact recovery, we evaluate contextual leakage by allowing the adversary to infer sensitive content from non-sensitive tokens. We adopt a SOTA approach~\cite{staab2023beyond} that uses observable tokens as context to reconstruct or approximate the sensitive content; detailed settings are in Appendix~\ref{sec:appendix_contextual_leakage}.

\bheading{Results (Table~\ref{tab:attack}).}
We observe that side-channel attacks cannot directly recover any token marked as sensitive. 
Beyond direct recovery, contextual leakage exists but remains limited: on average, only 2.2\% of sensitive tokens are exactly recovered and 6.4\% are semantically recovered. 
We further analyze the failed contextual-recovery cases and compare the attack requirements with the information exposed by \sysname, leading to the following takeaways.

\begin{table}[t]
\centering
\caption{Defense effectiveness of \sysname.}
\label{tab:attack}
\begin{tabular}{lccc}
\toprule
\multirow{2}{*}{\textbf{Dataset}} & \multirow{2}{*}{\textbf{Direct Recovery}} & \multicolumn{2}{c}{\textbf{Contextual Leakage}} \\
\cmidrule(lr){3-4}
 &  & Exact & Semantic Similar \\
\midrule
QASPER       & 0\% & 3.21\% & 9.48\% \\
NarrativeQA  & 0\% & 2.90\% & 5.88\% \\
QMSum        & 0\% & 0.50\% & 3.98\% \\
\bottomrule
\end{tabular}
\end{table}

\begin{packeditemize}
    \item \textit{Sensitive-token positions and counts are not identifiable.}
    The absence of reuse signals does not uniquely correspond to sensitive tokens, as not all non-sensitive tokens are shared in practice. After removing sensitive tokens, only a subset of non-sensitive content is further selected as reusable segments through self-contextualization. As a result, the set of non-shared tokens strictly exceeds the set of sensitive tokens, preventing the adversary from inferring the exact positions or counts of sensitive tokens. 
    \item \textit{Prompt-level structure is not reconstructible from reusable segments.}
    Reusable segments are stored independently without maintaining prompt-level grouping or ordering information. Even if an adversary recovers all reusable segments, it cannot reconstruct their original associations or sequence within the prompt. 
\end{packeditemize}

\begin{figure}[t]
    \centering
    \begin{minipage}{0.23\textwidth}
        \centering
        \includegraphics[width=\linewidth]{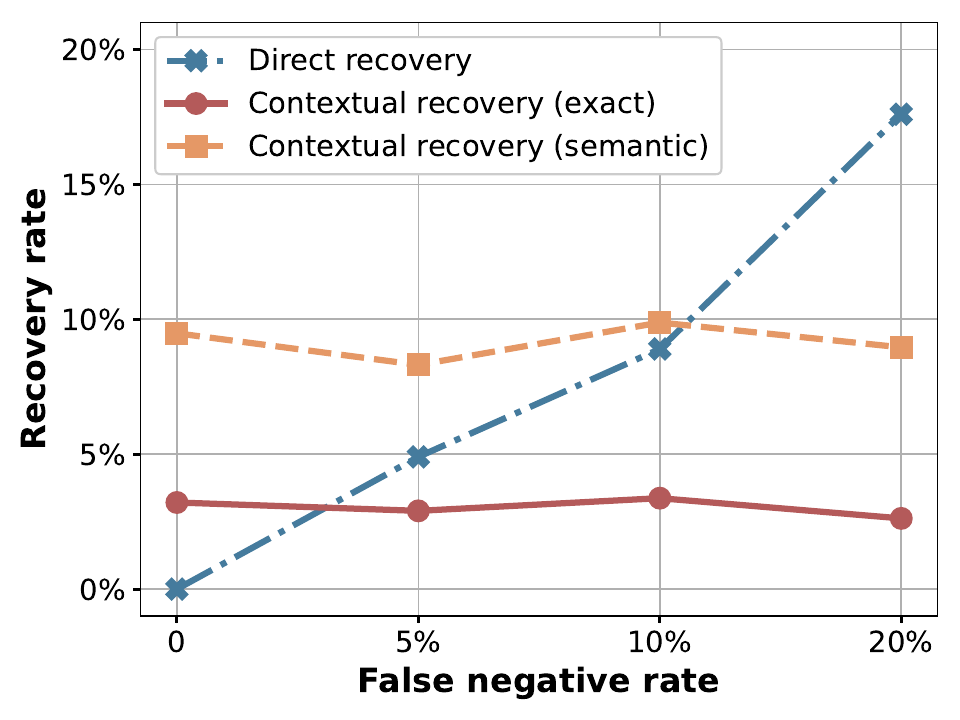}
        {\scriptsize (a) False negative}
        \label{fig:detecterror-a}
    \end{minipage}
    \hfill
    \begin{minipage}{0.23\textwidth}
        \centering
        \includegraphics[width=\linewidth]{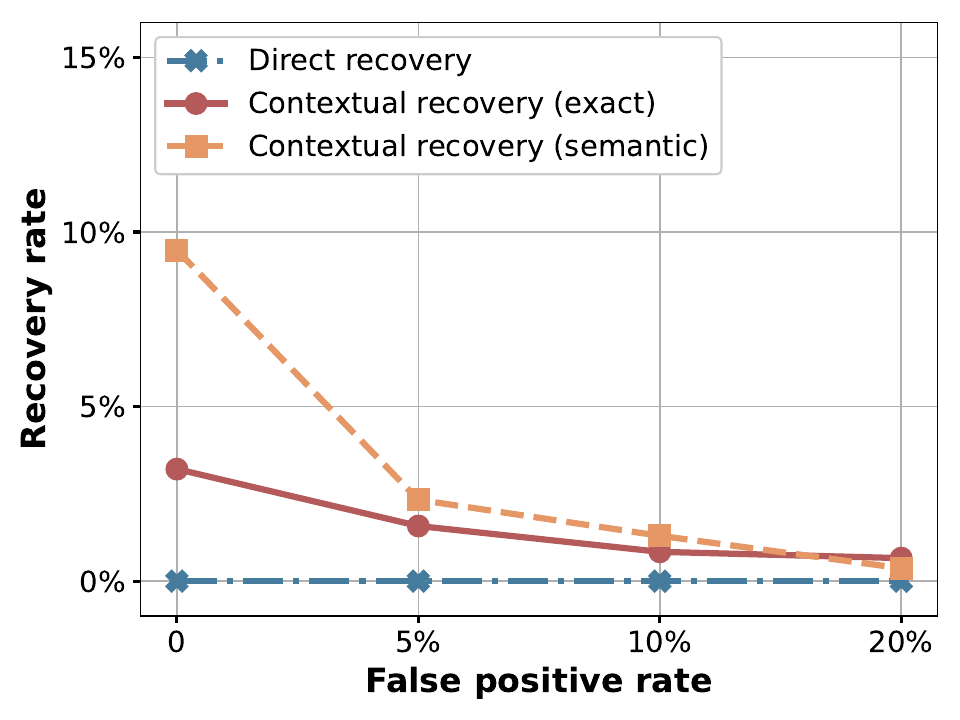}
        {\scriptsize (b) False positive}
        \label{fig:detecterror-b}
    \end{minipage}

    \caption{Impact of imperfect privacy detection (QASPER).}
    \label{fig:detecterror}
\end{figure}

\bheading{Impact of imperfect detection (\figref{fig:detecterror}).}
Privacy detection may introduce mislabeling, but its accuracy is difficult to quantify due to the lack of ground-truth PII annotations.
To assess its impact, we inject errors by simulating 0–20\% \textbf{F}alse \textbf{N}egatives (FN) and \textbf{F}alse \textbf{P}ositives (FP), covering and exceeding typical reported rates (both below 10\% for Presidio~\cite{presidio}). 
We randomly perturb the sensitivity mask and evaluate on all datasets, presenting QASPER in the main text and others in Appendix~\ref{sec:appendix_contextual_leakage}.
FN directly increase exact recovery by exposing unmarked sensitive tokens, while having negligible impact on contextual leakage. In contrast, FP do not affect direct recovery but significantly reduce contextual leakage by limiting reusable context, lowering the success rate of contextual inference.
Overall, detection quality mainly affects direct leakage through FN, while FP act conservatively by suppressing contextual leakage, motivating the use of conservative detectors.

\bheading{Summary.}
Tokens marked as sensitive cannot be recovered through reuse-based side channels. 
Contextual leakage remains limited, as \sysname does not expose structural information needed for inference. 
Imperfect detection mainly affects direct recovery, while having negligible impact on contextual leakage.

\subsection{RQ2: Generation Quality}
\label{subsec:rq1}

In \sysname, KV cache reuse is governed by three key factors—match rate, recompute rate, and segment length—which together determine generation quality.

\begin{packeditemize}

\item \textit{Match rate} refers to how much of the input prompt can be matched with previously stored KV segments, and a higher match rate means more of the prompt context is reused. 
\item \textit{Recompute rate} captures the proportion of KV entries within reused segments that is recomputed to preserve correctness in a new context. 
A lower rate means more direct reuse, while a higher rate indicates heavier recomputation. 
\item \textit{Segment length} denotes the size of the reused KV segments. 
Longer segments preserve more contextual information, which may help the model generate coherent outputs.
\end{packeditemize}
All figures in this section use Mistral-7B; Appendix~\ref{sec:appendix_generation_quality} reports results on Qwen-7B and Qwen-14B, which exhibit consistent results.

\begin{figure*}[!t]
\begin{minipage}{0.32\textwidth}
 	\centering
    \includegraphics[width=\textwidth]{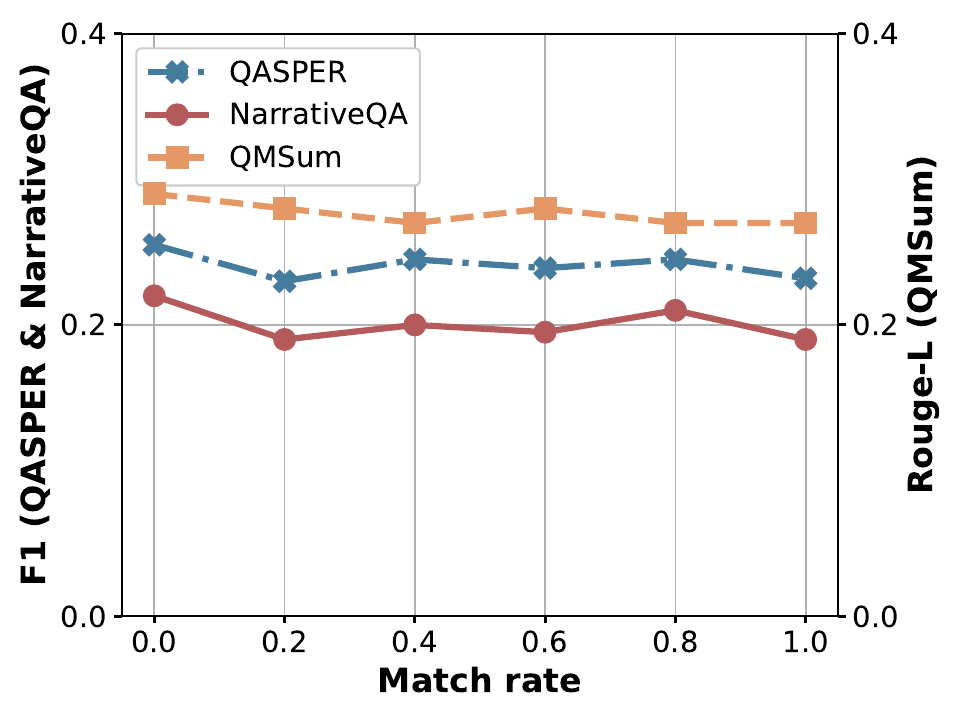}  
    \caption{Match rate impact (Mistral-7B).} \label{fig:matchrateimpact}
\end{minipage}
\begin{minipage}{0.1\textwidth}
\end{minipage}
\begin{minipage}{0.32\textwidth} 
    \includegraphics[width=\textwidth]{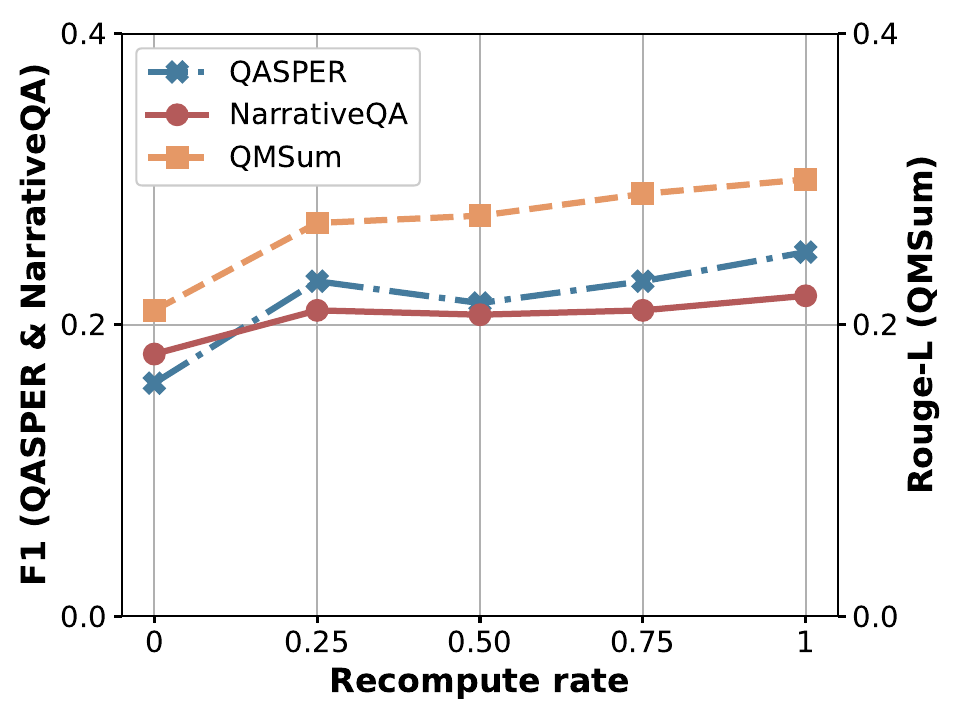}  
    \caption{Recompute rate impact (Mistral-7B).} \label{fig:recomputeimpact}
\end{minipage}
\begin{minipage}{0.1\textwidth}
\end{minipage}
\begin{minipage}{0.32\textwidth} 
    \includegraphics[width=\textwidth]{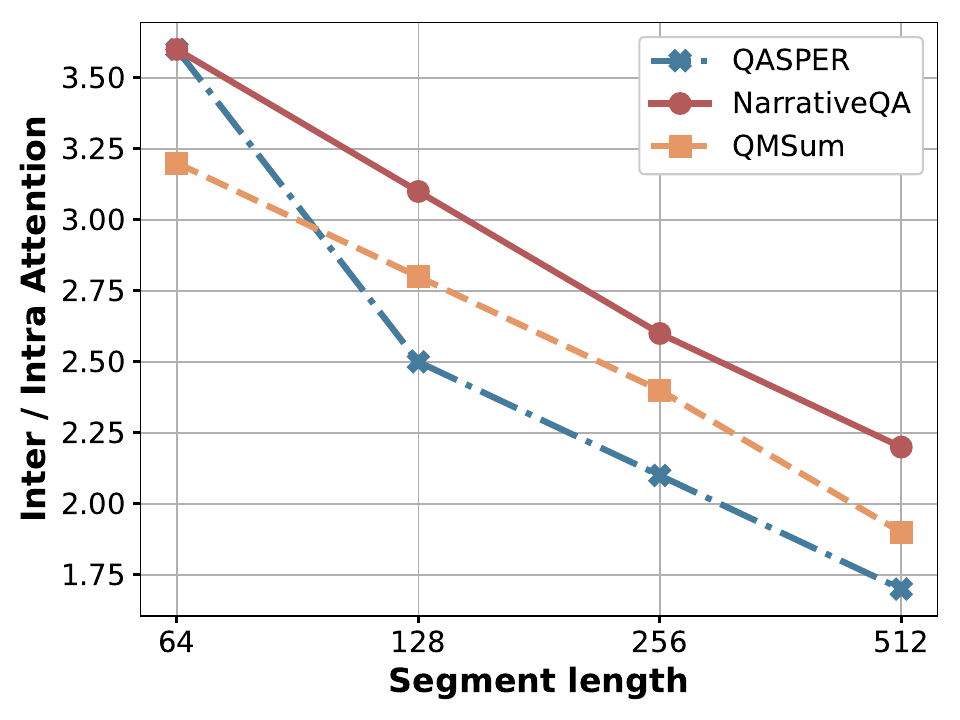}  
     \caption{Segment length impact (Mistral-7B).} \label{fig:segmentimpact}
\end{minipage}
\end{figure*} 

\subsubsection{Impact of Match Rate}
\label{subsubec:match}
We next examine how the match rate influences the generation quality.

\bheading{Methodology.}
We randomly sample 200 question–passage pairs from each dataset. 
For each passage $p$, we construct a related but distinct passage $p'$ using the LLM and form $p' + p$. 
We compute the KV cache of $p$ in this perturbed context to obtain reusable segments. 
We vary the match rate by controlling the fraction of reused KV cache from 0\% to 100\%, recomputing the remainder. 
Throughout this experiment, we fix the recompute rate at 25\%, which serves as our default setting and will be further examined in \secref{subsubsec:recompute}.

\bheading{Results (\figref{fig:matchrateimpact}).}
A match rate of 0\% corresponds to full recomputation (baseline), while 100\% represents full reuse (worst case). 
Across this range, accuracy remains stable on all datasets, indicating that match rate has minimal impact on generation quality in \sysname. 
This suggests that cache matches can be freely reused without degrading generation quality.

\subsubsection{Impact of Recompute Rate}
\label{subsubsec:recompute}
We next evaluate how recompute rate affects generation quality.

\bheading{Methodology.}
We follow the same setup as in \secref{subsubec:match}. 
We vary the recompute rate by selecting the top-ranked tokens based on intra- and inter-attention scores for recomputation, sweeping from 0\% to 100\% (including 25\%, 50\%, and 75\%).

\begin{figure}[t]
    \centering
    \begin{minipage}{0.23\textwidth}
        \centering
        \includegraphics[width=\linewidth]{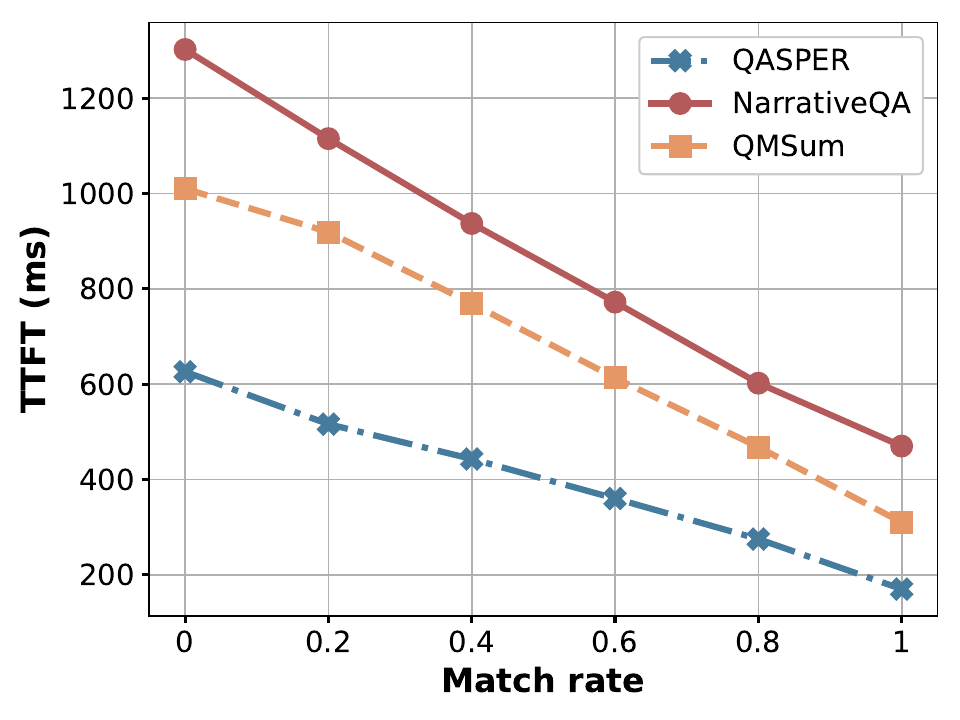}
        {\scriptsize (a) TTFT}
        \label{fig:match-a}
    \end{minipage}
    \hfill
    \begin{minipage}{0.23\textwidth}
        \centering
        \includegraphics[width=\linewidth]{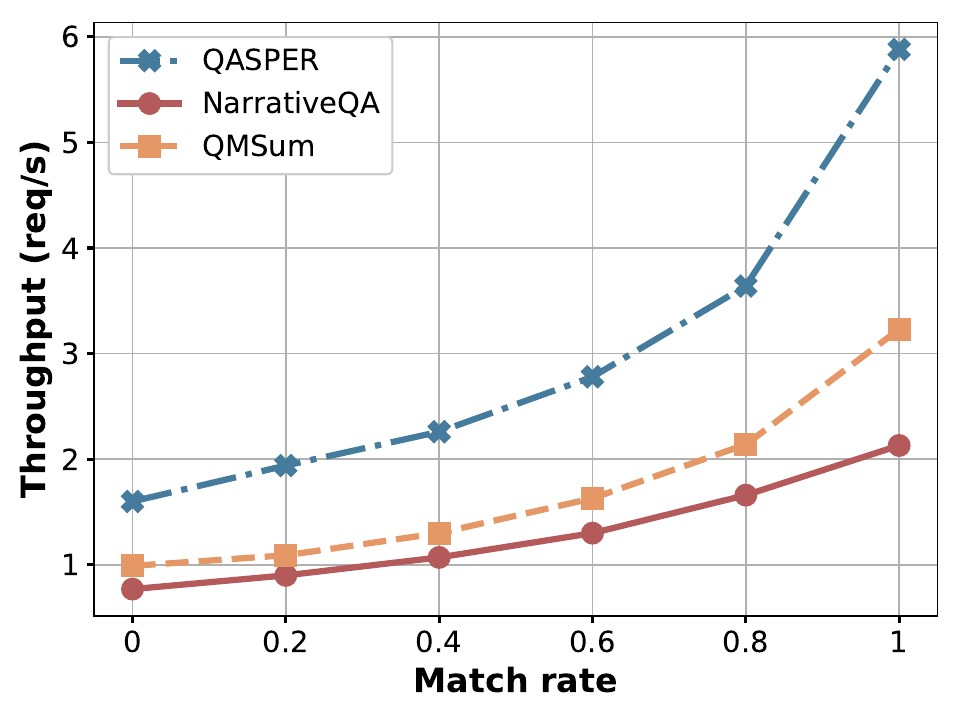}
        {\scriptsize (b) Throughput}
        \label{fig:match-b}
    \end{minipage}

    \caption{Impact of match rate on efficiency.}
    \label{fig:matchimpacteff}
\end{figure}

\bheading{Results (\figref{fig:recomputeimpact}).}
At 0\% recompute, where the KV cache is reused without any correction, the accuracy drops noticeably. 
Once recomputation is introduced, the results quickly stabilize, and the curves remain relatively flat up to 100\% recompute. 
This trend is consistent with prior observations in existing work~\cite{agarwal2025cache,yao2025cacheblend}. 
Based on this, we set 25\% recompute as the default rate in \sysname.

\subsubsection{Impact of Segment Length}
\label{subsubsec:length}
We next evaluate how segment length affects generation quality.

\bheading{Methodology.}
Longer segments are expected to preserve more contextual information as documented in prior work~\cite{agarwal2025cache,yao2025cacheblend}. 
We divide each request into fixed-size chunks and compute their inter- and intra-attention scores, where a higher inter-attention score indicates stronger dependence on outside context and thus lower reusability. 
We evaluate four chunk sizes, ranging from 64 to 512 tokens, using 200 randomly selected requests from each dataset.

\bheading{Results (\figref{fig:segmentimpact}).}
Across all datasets, the inter/intra ratio decreases as segment length increases, showing that longer segments are more self-contained and thus more reusable. 
Note that fixed-size chunks are used here only for analysis; in \sysname, segments are dynamically determined by context and are not directly comparable. 
In practice, we use 128 tokens as the minimum segment length, which can be adjusted as a system parameter.

\begin{figure}[t]
    \centering
    \begin{minipage}{0.23\textwidth}
        \centering
        \includegraphics[width=\linewidth]{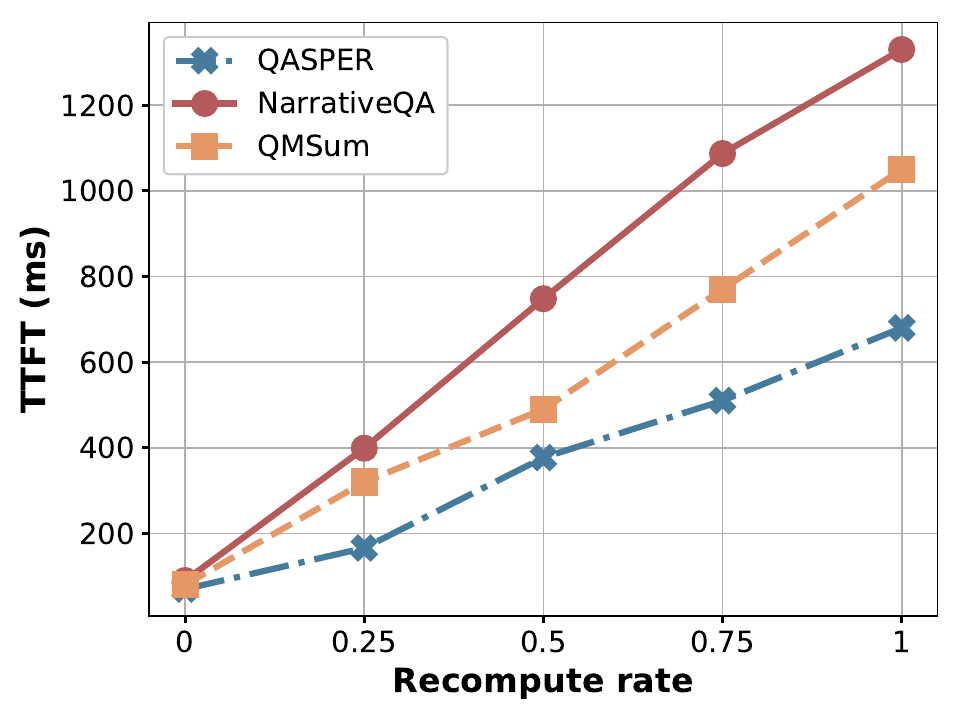}
        {\scriptsize (a) TTFT}
    \end{minipage}
    \hfill
    \begin{minipage}{0.23\textwidth}
        \centering
        \includegraphics[width=\linewidth]{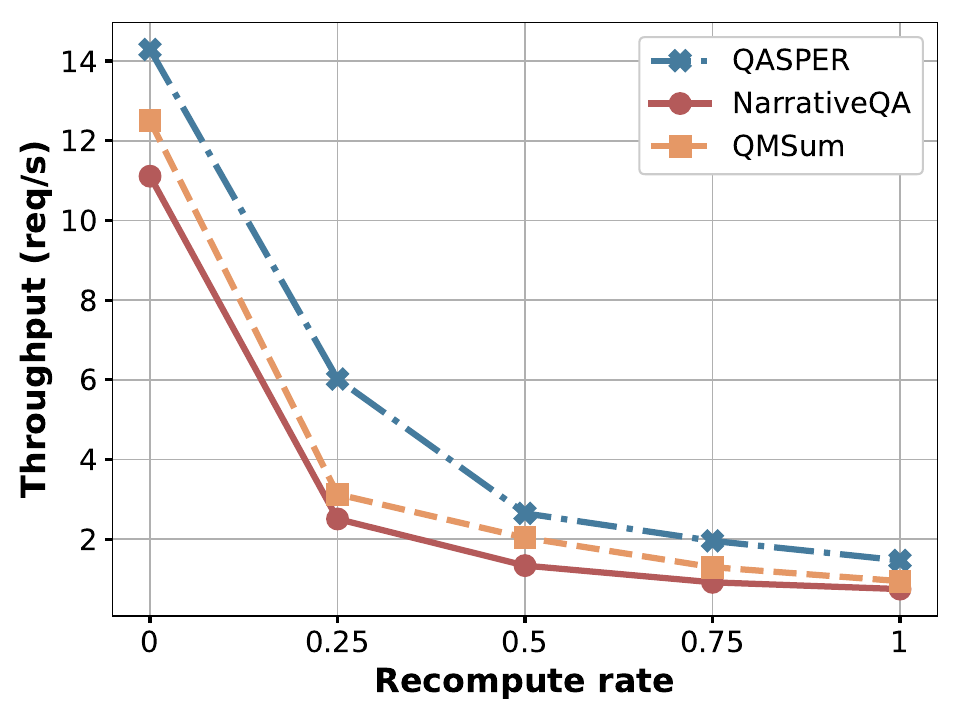}
        {\scriptsize (b) Throughput}
    \end{minipage}

    \caption{Impact of recompute rate on efficiency.}
    \label{fig:recomputeimpacteff}
\end{figure}

\begin{figure*}[!t]
    \centering
    \begin{minipage}{0.30\textwidth}
        \centering
        \includegraphics[width=\linewidth]{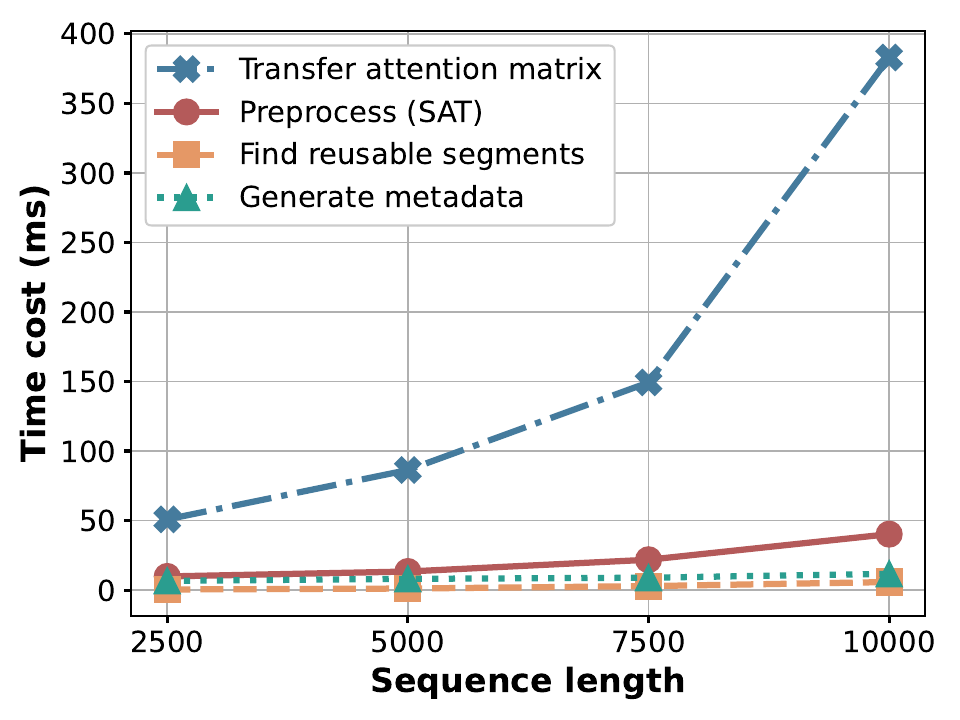}
        {\scriptsize (a)~\texttt{KV Annotator.}}
        \label{fig:annotatorcost}
    \end{minipage}
    \hfill
    \begin{minipage}{0.30\textwidth}
        \centering
        \includegraphics[width=\linewidth]{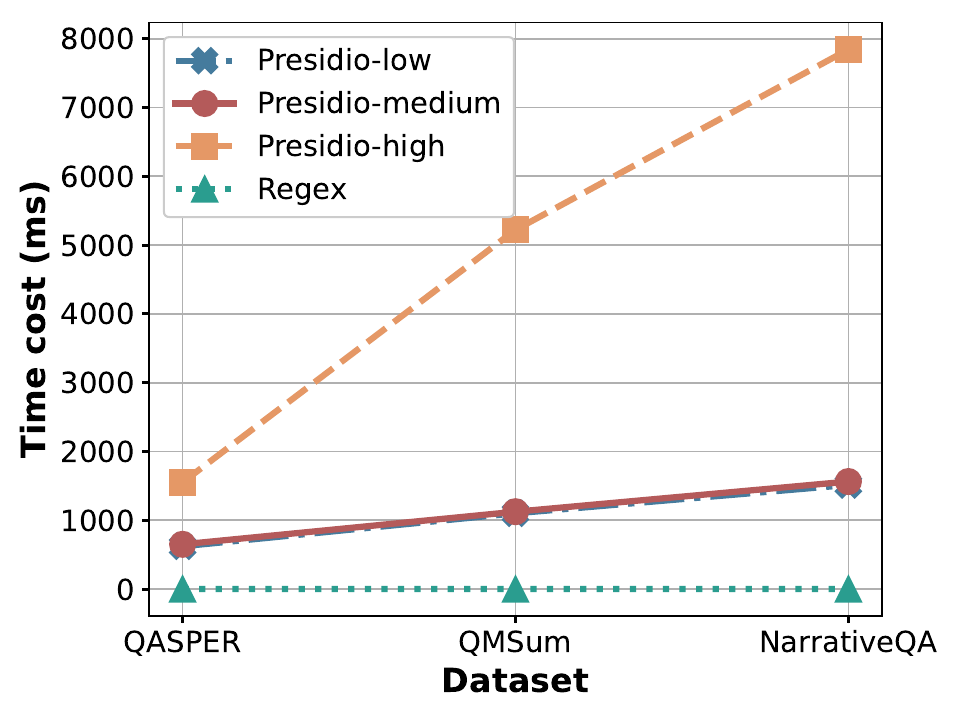}
        {\scriptsize (b) ~\texttt{Sensitivity Detector.}}
        \label{fig:detectorcost}
    \end{minipage}
    \hfill
    \begin{minipage}{0.30\textwidth}
        \centering
        \includegraphics[width=\linewidth]{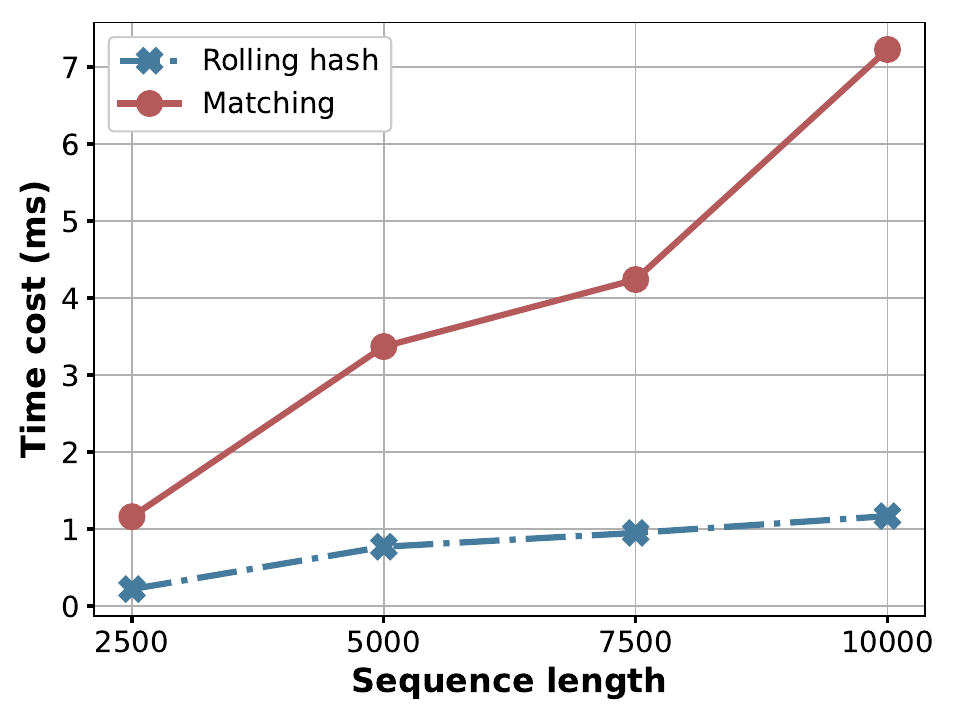}
        {\scriptsize (c) ~\texttt{KV Retriever.}}
        \label{fig:retrievercost}
    \end{minipage}

    \caption{Computation overhead of core components in \sysname.}
    \label{fig:componentcost}
\end{figure*}

\subsection{RQ3: Efficiency}
\label{subsec:rq2}

\subsubsection{Determining Factors}
The efficiency of \sysname is governed by the same three factors as generation quality: match rate, recompute rate, and segment length. 
We follow the setup in Sec.~\ref{subsec:rq1} and evaluate efficiency using TTFT and throughput. 
TTFT is measured over 200 independent requests as the average response latency, and throughput is measured as the average completed requests per second under a fixed arrival rate of one request per second over 30 seconds. 
All results are based on Mistral-7B; Appendix~\ref{sec:appendix_efficiency} reports consistent results on Qwen-7B and Qwen-14B.

\bheading{Impact of match rate (\figref{fig:matchimpacteff}).}
As the match rate increases, efficiency improves almost linearly, with TTFT reduced by up to 3$\times$ compared to the no-sharing baseline. 
This improvement stems from the greater proportion of KV being reused, reducing redundant computation. 
Although the absolute values differ across datasets due to variations in prompt length, the trend remains consistent.

\bheading{Impact of recompute rate (\figref{fig:recomputeimpacteff}).}
TTFT increases linearly with the recompute rate, as more tokens require fresh KV computation. 
At the default setting of 25\%, \sysname achieves over 3$\times$ lower TTFT than the non-sharing baseline (100\% recompute).

\bheading{Impact of segment length (\figref{fig:lengthimpacteff} in Appendix~\ref{sec:appendix_efficiency}).}
With a fixed recompute rate, smaller and more fragmented reusable segments incur additional overhead for linking and managing KV entries. 
As segment length increases, this overhead diminishes since longer segments reduce metadata lookups and cache-boundary operations. 
However, the overall difference remains marginal, suggesting \sysname maintains stable across different granularities.

\subsubsection{Detailed Breakdown of System Overhead}
\label{subsubsec:breakdown}
To further understand where the overhead of \sysname comes from, we break down the time and memory costs of system components, specifically \texttt{KV Retriever}, \texttt{Sensitivity Detector} and \texttt{KV Annotator}.

\bheading{Methodology.}
We measure the overhead of each component separately. 
For \texttt{KV Retriever} and \texttt{KV Annotator}, we vary the input length from 2,500 to 10,000 tokens in steps of 2,500. 
For the \texttt{Sensitivity Detector}, we evaluate on the three datasets using Presidio~\cite{presidio} under three privacy levels (low, medium, high), following existing methodology~\cite{ntwali2025detection} (details in Appendix~\ref{appendix:privacy}). 
We also include a regex-based baseline that masks randomly selected keywords. 
All results are averaged over 200 runs.

\begin{table*}[t]
\setlength{\tabcolsep}{9.5pt}
\centering
\caption{End-to-end evaluation results on three datasets with default settings.}
\label{tab:endtoendresults}
\begin{tabular}{lcccccccc}
\hline
\multirow{2}{*}{\textbf{Dataset}}
& \multirow{2}{*}{\textbf{Hit Rate}}
& \multicolumn{2}{c}{\textbf{TTFT (ms)}}
& \multicolumn{2}{c}{\textbf{Throughput (req/s)}}
& \multicolumn{2}{c}{\textbf{Generation Quality}}
\\ \cmidrule(lr){3-4} \cmidrule(lr){5-6} \cmidrule(lr){7-8}
&
&
\sysname & No-Sharing
& \sysname & No-Sharing
& \sysname & No-Sharing
\\
\hline
QASPER & 82.88\% & 273 & 797 & 3.66 & 1.25 & 0.28 & 0.30 \\
NarrativeQA & 56.45\% & 705 & 1675 & 1.42 & 0.60  & 0.23 & 0.25 \\
QMSum & 94.57\% & 342 & 1554 & 2.92 & 0.64 & 0.24 & 0.24 \\
\hline
\end{tabular}
\end{table*}

\bheading{KV Annotator (\figref{fig:componentcost}-a).}
The \texttt{KV Annotator} consists of four steps: (1) transferring the attention matrix from GPU to CPU, (2) building the summed area table, (3) identifying reusable segments, and (4) generating metadata (e.g., hashes and indices). 
The dominant overhead comes from data transfer, which takes up to 383\,ms for a 10,000-token request. 
Metadata generation and indexing contribute up to 50\,ms, while the other steps remain below 10\,ms each. 
Overall, the total cost is about 450\,ms for a 10,000-token request. 
Since the annotator runs after the response is returned and operates entirely on the CPU, it has limited impact on TTFT. 
We therefore do not optimize this component further and leave reducing transfer overhead (\eg, via pipelining) and accelerating summed-area table construction (\eg, via GPU parallelization~\cite{harris2007parallel}) to future work.

\bheading{Sensitivity Detector (\figref{fig:componentcost}-b).}
Different privacy detection levels incur substantially different costs. 
The strictest setting reaches up to 8s on NarrativeQA ($\approx$30K tokens), while a lower-privacy configuration incurs up to 600ms; in contrast, the regex-based baseline adds only 0.2ms. 
This shows that detection cost is dominated by the underlying implementation. 
As detection runs after response generation, it does not affect per-request latency such as TTFT. 
We therefore do not optimize this component further.

\bheading{KV Retriever (\figref{fig:componentcost}-c).}
The \texttt{KV Retriever} performs two operations: rolling-hash computation and two-phase matching. 
As it runs before each request, it represents the unavoidable overhead of \sysname. 
However, the total cost is only 8.3\,ms for a 10,000-token request, with 7.2\,ms spent on matching and 1.1\,ms on hashing. 
Compared to a TTFT of around 1,000\,ms, this overhead is negligible.

\subsection{RQ4: Impact of Privacy Detection}
\label{subsec:rq3}

The previous sections have shown that match rate, recompute rate, and segment length determine both generation quality and efficiency. 
Building on this, we observe that privacy detection policies directly affect the effective match rate and segment length, and thereby influence overall quality and efficiency, whereas the recompute rate is fixed by system design and unaffected by policies.

\bheading{Methodology.}
We follow the setup in \secref{subsubsec:breakdown} and evaluate \sysname under different privacy detection policies. 
We vary the detector from low to high privacy levels and measure the resulting changes in match rate and segment length, along with their impact on generation quality and efficiency.

\bheading{Impact on match rate (\figref{fig:privacyimpacteff}-a).} 
Stricter privacy policies reduce the match rate by marking more tokens as sensitive. However, since sensitive tokens constitute only a small portion of prompts, over 95\% of segments remain reusable even under the strictest setting.

\bheading{Impact on segment length (\figref{fig:privacyimpacteff}-b).}
Stricter privacy policies lead to shorter reusable segments, as more tokens are marked as sensitive and split prompts into smaller pieces.

\subsection{End-to-End Results} 
\label{subsec:endtoend}

We now evaluate \sysname under full end-to-end settings to capture its practical effectiveness.

\bheading{Methodology.}
For each dataset, we apply the default privacy detector to identify sensitive tokens and replace them with contextually consistent substitutes (generated by LLM), yielding a constructed dataset. 
We randomly sample 200 examples per dataset. 
For each example, we first issue the constructed request to populate reusable KV segments, followed by the original request to evaluate reuse. 
Requests are sent at a fixed rate of one per second, and results are averaged over all examples. 
We use Mistral-7B as the representative model, as prior results show consistent trends across models.

\bheading{Results (\tabref{tab:endtoendresults}).}
\sysname maintains generation quality comparable to no-sharing while achieving significant efficiency gains. 
In particular, TTFT improves by up to 4.5$\times$, with upmost 94.57\% of KV being reused. 
NarrativeQA shows a relatively lower hit rate, which we attribute to its data containing a larger fraction of privacy-sensitive tokens. 
This leads to more fragmented segments that cannot be reused, consistent with our observations in \secref{subsec:rq3}.

\subsection{Improvement over Fixed-Chunk Reuse}
\label{subsec:minorcontri}

\bheading{Methodology.}
We compare \sysname with SOTA fixed-chunk approaches~\cite{agarwal2025cache,yao2025cacheblend,hu2024epic} to evaluate the benefit of token-level reuse. 
To isolate the reuse mechanism, we disable privacy constraints in \sysname. 
Unlike the datasets used in earlier evaluations, which are constructed to provide ground-truth annotations for privacy analysis and thus exhibit limited natural overlap, here we use three real-world datasets (WildChat~\cite{wildchat}, LMSys~\cite{lmsys}, and ShareGPT~\cite{shareGPT}) to capture realistic reuse patterns; such datasets lack ground truth and are therefore unsuitable for the main evaluation but necessary for faithful efficiency analysis. 
Baselines partition each request into fixed-size chunks (128 tokens), while \sysname performs variable-length fine-grained reuse. 
All systems are evaluated under the same model (Mistral-7B) and serving configuration. 
We report the proportion of KV cache reused and the resulting efficiency.

\bheading{Results (Table~\ref{tab:kv_reuse_comparison}).}
\sysname significantly increases reuse opportunities over fixed-chunk methods, achieving up to 44\% higher match rates across datasets. 
This directly translates to lower TTFT, as more KV cache can be reused and less computation is required. 
Notably, match rate dominates system performance—once reuse is sufficiently high, differences among reuse strategies have limited impact compared to the gain from increased matching.

\begin{table}[t]
\setlength{\tabcolsep}{2.5pt}
\centering
\caption{Comparison of fine-grained and fixed-chunk KV reuse.}
\label{tab:kv_reuse_comparison}
\begin{tabular}{lcccc}
\toprule
\makecell[l]{\textbf{Dataset} \\ \quad Metrics} & 
\makecell[c]{\textbf{CachePrune} \\ (w/o privacy)} & 
\textbf{CacheCraft} & 
\textbf{CacheBlend} & 
\textbf{EPIC} \\
\midrule
\multicolumn{5}{l}{\textbf{WildChat}} \\
\quad Match Rate & 50.2\% & 22.5\% & 22.7\% & 25.1\% \\
\quad TTFT (ms) & 134 & 178 & 171 & 165 \\
\midrule
\multicolumn{5}{l}{\textbf{ShareGPT}} \\
\quad Match Rate & 30.5\% & 7.73\% & 7.92\% & 8.57\% \\
\quad TTFT (ms) & 177 & 213 & 210 & 209 \\
\midrule
\multicolumn{5}{l}{\textbf{LMSys}} \\
\quad Match Rate & 47.3\% & 2.9\% & 3.1\% & 3.1\% \\
\quad TTFT (ms) & 74 & 98 & 93 & 94 \\
\bottomrule
\end{tabular}
\end{table}

\section{Related Work}
\label{sec:related}

\bheading{Selective KV handling.}
Recent work explores selective KV handling to improve inference efficiency, including KV dropping~\cite{tang2024razorattention,wang2025llms}, offloading~\cite{lee2024infinigen,luo2025headinfer}, and quantization~\cite{hooper2024kvquant,su2025accurate}. 
These approaches differentiate tokens based on their contribution to output quality. 
In contrast, our work follows a similar selective paradigm but selects based on sensitivity, shifting the focus from efficiency to privacy.

\bheading{Defenses against side channel attacks on shared resources.}
Shared resources introduce security risks~\cite{wucache}, especially side channels that leak information through shared state.
Existing defenses fall into three categories: partitioning~\cite{dong2018shielding,liu2016catalyst}, randomization~\cite{werner2019scattercache,ramkrishnan2019new}, and obfuscation~\cite{nam2023defensive}. 
Partitioning eliminates sharing by isolating resources (e.g., cache coloring), while randomization and obfuscation preserve sharing but reduce leakage via unpredictability or noise. 
Among these, partitioning provides the strongest protection, motivating our design to disable sharing for sensitive tokens.

\section{Discussion and Future Work}
\label{sec:discussion}

\bheading{Decoupled privacy detection.}
Privacy detection is orthogonal to our design: \sysname provides a unified interface that supports different detections.
Privacy detection is a well-established research area that is widely adopted in both academic~\cite{nadeau2007survey} and production~\cite{azurepii,openaiprivacyfilter}, and has been further strengthened by recent LLM-based techniques~\cite{ponomarenko2026capid,openaiprivacyfilter}, making it a practical security-by-design foundation. 
We acknowledge privacy detection is not perfect and we evaluate the impact of such imperfections in \secref{subsec:rq0}, leaving improved detection to future work.

\bheading{Scope of KV cache security.}
KV cache has become a critical component in modern LLM serving, and its security directly impacts both user inputs and model weights. 
Prior work has shown that KV reuse introduces practical risks such as side-channel leakage~\cite{wu2025know}, while other emerging settings consider stronger assumptions where KV entries may be exposed in plaintext and inverted through transformer structure~\cite{luo2025shadow}. 
In this paper, we focus on side-channel attacks, as they represent a practical threat in real-world multi-tenant systems where KV cache remains internal and not directly observable. We do not consider scenarios where KV is explicitly exposed and leave such stronger threat models for future work.

\section{Conclusion}
\label{sec:conclusion}

We introduced \sysname, a privacy-aware and fine-grained KV cache sharing system that enables secure and efficient reuse of KV states in multi-tenant LLM serving. By moving beyond fixed-size chunking and supporting token-level reuse, \sysname derives and retrieves reusable segments with low overhead while excluding sensitive content from sharing. Our prototype on vLLM shows substantial gains over existing system, demonstrating that fine-grained KV management is both practical and provides a foundation for broader future optimizations in LLM serving.

\normalem
\bibliographystyle{ACM-Reference-Format}
\bibliography{bib}

@article{zhao2023survey,
  title={A survey of large language models},
  author={Zhao, Wayne Xin and Zhou, Kun and Li, Junyi and Tang, Tianyi and Wang, Xiaolei and Hou, Yupeng and Min, Yingqian and Zhang, Beichen and Zhang, Junjie and Dong, Zican and others},
  journal={arXiv preprint arXiv:2303.18223},
  volume={1},
  number={2},
  year={2023}
}

@article{naveed2023comprehensive,
  title={A comprehensive overview of large language models},
  author={Naveed, Humza and Khan, Asad Ullah and Qiu, Shi and Saqib, Muhammad and Anwar, Saeed and Usman, Muhammad and Akhtar, Naveed and Barnes, Nick and Mian, Ajmal},
  journal={ACM Transactions on Intelligent Systems and Technology},
  year={2023},
  publisher={ACM New York, NY}
}

@article{harris2007parallel,
  title={Parallel prefix sum (scan) with CUDA},
  author={Harris, Mark and Sengupta, Shubhabrata and Owens, John D},
  journal={GPU gems},
  volume={3},
  number={39},
  pages={851--876},
  year={2007}
}

@inproceedings{yan2024protecting,
  title={On protecting the data privacy of large language models (llms): A survey},
  author={Yan, Biwei and Li, Kun and Xu, Minghui and Dong, Yueyan and Zhang, Yue and Ren, Zhaochun and Cheng, Xiuzhen},
  booktitle={2024 International Conference on Meta Computing (ICMC)},
  pages={1--12},
  year={2024},
  organization={IEEE}
}

@inproceedings{ponomarenko2026capid,
  title={CAPID: Context-Aware PII Detection for Question-Answering Systems},
  author={Ponomarenko, Mariia and Abedini, Sepideh and Shafieinejad, Masoumeh and Emerson, DB and Mohapatra, Shubhankar and He, Xi},
  booktitle={Proceedings of the 19th Conference of the European Chapter of the Association for Computational Linguistics (Volume 4: Student Research Workshop)},
  pages={320--331},
  year={2026}
}

@article{vaswani2017attention,
  title={Attention is all you need},
  author={Vaswani, Ashish and Shazeer, Noam and Parmar, Niki and Uszkoreit, Jakob and Jones, Llion and Gomez, Aidan N and Kaiser, {\L}ukasz and Polosukhin, Illia},
  journal={Advances in neural information processing systems},
  volume={30},
  year={2017}
}

@article{nadeau2007survey,
  title={A survey of named entity recognition and classification},
  author={Nadeau, David and Sekine, Satoshi},
  journal={Lingvisticae Investigationes},
  volume={30},
  number={1},
  pages={3--26},
  year={2007},
  publisher={John Benjamins}
}

@article{staab2023beyond,
  title={Beyond memorization: Violating privacy via inference with large language models},
  author={Staab, Robin and Vero, Mark and Balunovi{\'c}, Mislav and Vechev, Martin},
  journal={arXiv preprint arXiv:2310.07298},
  year={2023}
}

@article{wang2026optileak,
  title={OptiLeak: Efficient Prompt Reconstruction via Reinforcement Learning in Multi-tenant LLM Services},
  author={Wang, Longxiang and Zheng, Xiang and Zhang, Xuhao and Zhang, Yao and Wu, Ye and Wang, Cong},
  journal={arXiv preprint arXiv:2602.20595},
  year={2026}
}

@article{li2024survey,
  title={A survey on large language model acceleration based on kv cache management},
  author={Li, Haoyang and Li, Yiming and Tian, Anxin and Tang, Tianhao and Xu, Zhanchao and Chen, Xuejia and Hu, Nicole and Dong, Wei and Li, Qing and Chen, Lei},
  journal={arXiv preprint arXiv:2412.19442},
  year={2024}
}

@article{jin2024compute,
  title={Compute or load kv cache? why not both?},
  author={Jin, Shuowei and Liu, Xueshen and Zhang, Qingzhao and Mao, Z Morley},
  journal={arXiv preprint arXiv:2410.03065},
  year={2024}
}

@inproceedings{kwon2023efficient,
  title={Efficient memory management for large language model serving with pagedattention},
  author={Kwon, Woosuk and Li, Zhuohan and Zhuang, Siyuan and Sheng, Ying and Zheng, Lianmin and Yu, Cody Hao and Gonzalez, Joseph and Zhang, Hao and Stoica, Ion},
  booktitle={Proceedings of the 29th symposium on operating systems principles},
  pages={611--626},
  year={2023}
}

@article{zheng2024sglang,
  title={Sglang: Efficient execution of structured language model programs},
  author={Zheng, Lianmin and Yin, Liangsheng and Xie, Zhiqiang and Sun, Chuyue Livia and Huang, Jeff and Yu, Cody Hao and Cao, Shiyi and Kozyrakis, Christos and Stoica, Ion and Gonzalez, Joseph E and others},
  journal={Advances in neural information processing systems},
  volume={37},
  pages={62557--62583},
  year={2024}
}

@inproceedings{yao2025cacheblend,
  title={CacheBlend: Fast large language model serving for RAG with cached knowledge fusion},
  author={Yao, Jiayi and Li, Hanchen and Liu, Yuhan and Ray, Siddhant and Cheng, Yihua and Zhang, Qizheng and Du, Kuntai and Lu, Shan and Jiang, Junchen},
  booktitle={Proceedings of the Twentieth European Conference on Computer Systems},
  pages={94--109},
  year={2025}
}

@article{agarwal2025cache,
  title={Cache-craft: Managing chunk-caches for efficient retrieval-augmented generation},
  author={Agarwal, Shubham and Sundaresan, Sai and Mitra, Subrata and Mahapatra, Debabrata and Gupta, Archit and Sharma, Rounak and Kapu, Nirmal Joshua and Yu, Tong and Saini, Shiv},
  journal={Proceedings of the ACM on Management of Data},
  volume={3},
  number={3},
  pages={1--28},
  year={2025},
  publisher={ACM New York, NY, USA}
}

@article{hu2024epic,
  title={EPIC: Efficient Position-Independent Caching for Serving Large Language Models},
  author={Hu, Junhao and Huang, Wenrui and Wang, Weidong and Wang, Haoyi and Hu, Tiancheng and Zhang, Qin and Feng, Hao and Chen, Xusheng and Shan, Yizhou and Xie, Tao},
  journal={arXiv preprint arXiv:2410.15332},
  year={2024}
}

@article{liu2024deepseek,
  title={Deepseek-v3 technical report},
  author={Liu, Aixin and Feng, Bei and Xue, Bing and Wang, Bingxuan and Wu, Bochao and Lu, Chengda and Zhao, Chenggang and Deng, Chengqi and Zhang, Chenyu and Ruan, Chong and others},
  journal={arXiv preprint arXiv:2412.19437},
  year={2024}
}

@misc{msmarco,
  author = {{Microsoft}},
  title={microsoft/msmarco},
  howpublished={\url{https://huggingface.co/datasets/microsoft/ms_marco}},
  year={2024}
}

@misc{openaipromptcaching,
  author = {{OpenAI}},
  title={Prompt caching},
  howpublished={\url{https://platform.openai.com/docs/guides/prompt-caching}},
  year={2025}
}

@misc{presidio,
  author = {{Microsoft}},
  title={Presidio: Data Protection and De-identification SDK},
  howpublished={\url{https://microsoft.github.io/presidio/}},
  year={2025}
}

@misc{wildchat,
  author = {{AllenAI}},
  title={allenai/WildChat},
  howpublished={\url{https://huggingface.co/datasets/allenai/WildChat}},
  year={2025}
}

@misc{lmsys,
  author = {{LMSys}},
  title={lmsys/lmsys-chat-1m},
  howpublished={\url{https://huggingface.co/datasets/lmsys/lmsys-chat-1m}},
  year={2025}
}

@misc{shareGPT,
  author = {{RyokoAI}},
  title={RyokoAI/ShareGPT52K},
  howpublished={\url{https://huggingface.co/datasets/RyokoAI/ShareGPT52K}},
  year={2025}
}

@misc{qasper,
  author = {{Qasper}},
  title={allenai/qasper},
  howpublished={\url{https://huggingface.co/datasets/allenai/qasper}},
  year={2025}
}

@misc{mistral,
  author = {{MistralAI}},
  title={mistralai/Mistral-7B-v0.1},
  howpublished={\url{https://huggingface.co/mistralai/Mistral-7B-v0.1}},
  year={2025},
}

@misc{qwen7b,
  author = {{Qwen}},
  title={Qwen/Qwen2.5-7B},
  howpublished={\url{https://huggingface.co/Qwen/Qwen2.5-7B}},
  year={2025},
}

@misc{qwen14b,
  author = {{Qwen}},
  title={Qwen/Qwen2.5-14B},
  howpublished={\url{https://huggingface.co/Qwen/Qwen2.5-14B}},
  year={2025},
}

@misc{azurepii,
  author = {{Azure}},
  title={What is PII detection in Azure Language?},
  howpublished={\url{https://learn.microsoft.com/en-us/azure/ai-services/language-service/personally-identifiable-information/overview?tabs=text-pii}},
  year={2025},
}

@article{wucache,
  title={When Cache Poisoning Meets LLM Systems: Semantic Cache Poisoning and Its Countermeasures},
  author={Wu, Guanlong and Wang, Taojie and Zhang, Yao and Zhang, Zheng and Niu, Jianyu and Wu, Ye and Zhang, Yinqian},
  year={2026},
}

@article{luo2025shadow,
  title={Shadow in the cache: Unveiling and mitigating privacy risks of kv-cache in llm inference},
  author={Luo, Zhifan and Shao, Shuo and Zhang, Su and Zhou, Lijing and Hu, Yuke and Zhao, Chenxu and Liu, Zhihao and Qin, Zhan},
  journal={arXiv preprint arXiv:2508.09442},
  year={2025}
}

@misc{narrativeqa,
  author = {{Deepmind}},
  title={deepmind/narrativeqa},
  howpublished={\url{https://huggingface.co/datasets/deepmind/narrativeqa}},
  year={2025},
}

@misc{qmsum,
  author = {{Yale-LTLY}},
  title={QMSum},
  howpublished={\url{https://github.com/Yale-LILY/QMSum}},
  year={2025},
}

@misc{nvidiakv,
  author = {{Nvidia}},
  title={Structuring Applications to Secure the KV Cache},
  howpublished={\url{https://developer.nvidia.com/blog/structuring-applications-to-secure-the-kv-cache/}},
  year={2025},
}

@misc{claudepromptcaching,
  author = {{Claude}},
  title={Prompt caching},
  howpublished={\url{https://docs.anthropic.com/en/docs/build-with-claude/prompt-caching}},
  year={2025},
}

@misc{geminipromptcaching,
  author = {{Gemini}},
  title={Context Caching},
  howpublished={\url{https://ai.google.dev/gemini-api/docs/caching}},
  year={2025},
}

@misc{openaiprivacyfilter,
  author = {{OpenAI}},
  title={OpenAI Privacy Filter},
  howpublished={\url{https://huggingface.co/openai/privacy-filter}},
  year={2025},
}

@inproceedings{wu2025know,
  title={I know what you asked: Prompt leakage via kv-cache sharing in multi-tenant llm serving},
  author={Wu, Guanlong and Zhang, Zheng and Zhang, Yao and Wang, Weili and Niu, Jianyu and Wu, Ye and Zhang, Yinqian},
  booktitle={Proceedings of the 2025 Network and Distributed System Security (NDSS) Symposium. San Diego, CA, USA},
  year={2025}
}

@article{song2024early,
  title={The early bird catches the leak: Unveiling timing side channels in llm serving systems},
  author={Song, Linke and Pang, Zixuan and Wang, Wenhao and Wang, Zihao and Wang, XiaoFeng and Chen, Hongbo and Song, Wei and Jin, Yier and Meng, Dan and Hou, Rui},
  journal={arXiv preprint arXiv:2409.20002},
  year={2024}
}

@article{zheng2024inputsnatch,
  title={Inputsnatch: Stealing input in llm services via timing side-channel attacks},
  author={Zheng, Xinyao and Han, Husheng and Shi, Shangyi and Fang, Qiyan and Du, Zidong and Hu, Xing and Guo, Qi},
  journal={arXiv preprint arXiv:2411.18191},
  year={2024}
}

@inproceedings{pang2024cache,
  title={Cache partitioning for mitigating timing side-channel attacks in llm serving systems},
  author={Pang, Zixuan and Wang, Wenhao and Liao, Yong},
  booktitle={2024 6th International Conference on Frontier Technologies of Information and Computer (ICFTIC)},
  pages={1238--1245},
  year={2024},
  organization={IEEE}
}

@article{tang2024razorattention,
  title={Razorattention: Efficient kv cache compression through retrieval heads},
  author={Tang, Hanlin and Lin, Yang and Lin, Jing and Han, Qingsen and Hong, Shikuan and Yao, Yiwu and Wang, Gongyi},
  journal={arXiv preprint arXiv:2407.15891},
  year={2024}
}

@article{wang2025llms,
  title={LLMs Know What to Drop: Self-Attention Guided KV Cache Eviction for Efficient Long-Context Inference},
  author={Wang, Guangtao and Upasani, Shubhangi and Wu, Chen and Gandhi, Darshan and Li, Jonathan and Hu, Changran and Li, Bo and Thakker, Urmish},
  journal={arXiv preprint arXiv:2503.08879},
  year={2025}
}

@inproceedings{crow1984summed,
  title={Summed-area tables for texture mapping},
  author={Crow, Franklin C},
  booktitle={Proceedings of the 11th annual conference on Computer graphics and interactive techniques},
  pages={207--212},
  year={1984}
}

@inproceedings{hensley2005fast,
  title={Fast summed-area table generation and its applications},
  author={Hensley, Justin and Scheuermann, Thorsten and Coombe, Greg and Singh, Montek and Lastra, Anselmo},
  booktitle={Computer Graphics Forum},
  volume={24},
  number={3},
  pages={547--556},
  year={2005},
  organization={Amsterdam: North Holland, 1982-}
}

@article{chayapathi2021survey,
  title={Survey and comparison of string matching algorithms},
  author={Chayapathi, AR and Kumar, G Sunil and Swamy, Manjunath BE and Thriveni, J and Venugopal, KR},
  journal={Turkish Journal of Computer and Mathematics Education},
  volume={12},
  number={12},
  pages={1471--1491},
  year={2021},
  publisher={Ninety Nine Publication}
}

@article{jiang2020rolling,
  title={A rolling hash algorithm and the implementation to LZ4 data compression},
  author={Jiang, Hao and Lin, Sian-Jheng},
  journal={IEEE Access},
  volume={8},
  pages={35529--35534},
  year={2020},
  publisher={IEEE}
}

@incollection{mohit2014named,
  title={Named entity recognition},
  author={Mohit, Behrang},
  booktitle={Natural language processing of semitic languages},
  pages={221--245},
  year={2014},
  publisher={Springer}
}

@article{bhojanapalli2021leveraging,
  title={Leveraging redundancy in attention with reuse transformers},
  author={Bhojanapalli, Srinadh and Chakrabarti, Ayan and Veit, Andreas and Lukasik, Michal and Jain, Himanshu and Liu, Frederick and Chang, Yin-Wen and Kumar, Sanjiv},
  journal={arXiv preprint arXiv:2110.06821},
  year={2021}
}

@article{vig2019analyzing,
  title={Analyzing the structure of attention in a transformer language model},
  author={Vig, Jesse and Belinkov, Yonatan},
  journal={arXiv preprint arXiv:1906.04284},
  year={2019}
}

@article{ferrando2022measuring,
  title={Measuring the mixing of contextual information in the transformer},
  author={Ferrando, Javier and G{\'a}llego, Gerard I and Costa-Juss{\`a}, Marta R},
  journal={arXiv preprint arXiv:2203.04212},
  year={2022}
}

@inproceedings{lee2024infinigen,
  title={$\{$InfiniGen$\}$: Efficient generative inference of large language models with dynamic $\{$KV$\}$ cache management},
  author={Lee, Wonbeom and Lee, Jungi and Seo, Junghwan and Sim, Jaewoong},
  booktitle={18th USENIX Symposium on Operating Systems Design and Implementation (OSDI 24)},
  pages={155--172},
  year={2024}
}

@article{luo2025headinfer,
  title={Headinfer: Memory-efficient llm inference by head-wise offloading},
  author={Luo, Cheng and Cai, Zefan and Sun, Hanshi and Xiao, Jinqi and Yuan, Bo and Xiao, Wen and Hu, Junjie and Zhao, Jiawei and Chen, Beidi and Anandkumar, Anima},
  journal={arXiv preprint arXiv:2502.12574},
  year={2025}
}

@article{su2025accurate,
  title={Accurate kv cache quantization with outlier tokens tracing},
  author={Su, Yi and Zhou, Yuechi and Qiu, Quantong and Li, Juntao and Xia, Qingrong and Li, Ping and Duan, Xinyu and Wang, Zhefeng and Zhang, Min},
  journal={arXiv preprint arXiv:2505.10938},
  year={2025}
}

@article{hooper2024kvquant,
  title={Kvquant: Towards 10 million context length llm inference with kv cache quantization},
  author={Hooper, Coleman and Kim, Sehoon and Mohammadzadeh, Hiva and Mahoney, Michael W and Shao, Yakun S and Keutzer, Kurt and Gholami, Amir},
  journal={Advances in Neural Information Processing Systems},
  volume={37},
  pages={1270--1303},
  year={2024}
}

@inproceedings{dong2018shielding,
  title={Shielding Software From Privileged $\{$Side-Channel$\}$ Attacks},
  author={Dong, Xiaowan and Shen, Zhuojia and Criswell, John and Cox, Alan L and Dwarkadas, Sandhya},
  booktitle={27th USENIX Security Symposium (USENIX Security 18)},
  pages={1441--1458},
  year={2018}
}

@inproceedings{liu2016catalyst,
  title={Catalyst: Defeating last-level cache side channel attacks in cloud computing},
  author={Liu, Fangfei and Ge, Qian and Yarom, Yuval and Mckeen, Frank and Rozas, Carlos and Heiser, Gernot and Lee, Ruby B},
  booktitle={2016 IEEE international symposium on high performance computer architecture (HPCA)},
  pages={406--418},
  year={2016},
  organization={IEEE}
}

@inproceedings{werner2019scattercache,
  title={$\{$ScatterCache$\}$: thwarting cache attacks via cache set randomization},
  author={Werner, Mario and Unterluggauer, Thomas and Giner, Lukas and Schwarz, Michael and Gruss, Daniel and Mangard, Stefan},
  booktitle={28th USENIX Security Symposium (USENIX Security 19)},
  pages={675--692},
  year={2019}
}

@article{ramkrishnan2019new,
  title={New attacks and defenses for randomized caches},
  author={Ramkrishnan, Kartik and Zhai, Antonia and McCamant, Stephen and Yew, Pen Chung},
  journal={arXiv preprint arXiv:1909.12302},
  year={2019}
}

@article{nam2023defensive,
  title={Defensive ml: Defending architectural side-channels with adversarial obfuscation},
  author={Nam, Hyoungwook and Pothukuchi, Raghavendra Pradyumna and Li, Bo and Kim, Nam Sung and Torrellas, Josep},
  journal={arXiv preprint arXiv:2302.01474},
  year={2023}
}

@article{touvron2023llama,
  title={Llama: Open and efficient foundation language models},
  author={Touvron, Hugo and Lavril, Thibaut and Izacard, Gautier and Martinet, Xavier and Lachaux, Marie-Anne and Lacroix, Timoth{\'e}e and Rozi{\`e}re, Baptiste and Goyal, Naman and Hambro, Eric and Azhar, Faisal and others},
  journal={arXiv preprint arXiv:2302.13971},
  year={2023}
}

@article{achiam2023gpt,
  title={Gpt-4 technical report},
  author={Achiam, Josh and Adler, Steven and Agarwal, Sandhini and Ahmad, Lama and Akkaya, Ilge and Aleman, Florencia Leoni and Almeida, Diogo and Altenschmidt, Janko and Altman, Sam and Anadkat, Shyamal and others},
  journal={arXiv preprint arXiv:2303.08774},
  year={2023}
}

@article{ntwali2025detection,
  title={Detection of Personal Data in Structured Datasets Using a Large Language Model},
  author={Ntwali, Albert Agisha and R{\"u}ck, Luca and Heckmann, Martin},
  journal={arXiv preprint arXiv:2506.22305},
  year={2025}
}

@article{schwartz2011pii,
  title={The PII problem: Privacy and a new concept of personally identifiable information},
  author={Schwartz, Paul M and Solove, Daniel J},
  journal={NYUL rev.},
  volume={86},
  pages={1814},
  year={2011},
  publisher={HeinOnline}
}

@article{narayanan2010myths,
  title={Myths and fallacies of" personally identifiable information"},
  author={Narayanan, Arvind and Shmatikov, Vitaly},
  journal={Communications of the ACM},
  volume={53},
  number={6},
  pages={24--26},
  year={2010},
  publisher={ACM New York, NY, USA}
}

@book{mccallister2010guide,
  title={Guide to protecting the confidentiality of personally identifiable information},
  author={McCallister, Erika},
  year={2010},
  publisher={Diane Publishing}
}

@article{ackerman2001privacy,
  title={Privacy in context},
  author={Ackerman, Mark and Darrell, Trevor and Weitzner, Daniel J},
  journal={Human--Computer Interaction},
  volume={16},
  number={2-4},
  pages={167--176},
  year={2001},
  publisher={Taylor \& Francis}
}

@article{nissenbaum2018respecting,
  title={Respecting context to protect privacy: Why meaning matters},
  author={Nissenbaum, Helen},
  journal={Science and engineering ethics},
  volume={24},
  number={3},
  pages={831--852},
  year={2018},
  publisher={Springer}
}
\appendix

\section{Defense effectiveness evaluations}
\label{sec:appendix_contextual_leakage}

\subsection{Attack Settings of Contextual Leakage}
We evaluate contextual leakage by measuring whether sensitive information can be inferred from non-sensitive tokens after anonymization. The attack follows a guess–judge pipeline as in prior work~\cite{staab2023beyond}.

\bheading{Data preprocessing.}
We first identify and anonymize all PII using Presidio with spaCy on GPU. Each detected entity is replaced with a string of '*' of the same length, ensuring that sensitive tokens are fully masked while preserving surrounding context.

\bheading{Guesser.}
For each sample, we group PII instances into batches of 100 and prompt a guesser LLM to infer the masked content. For each instance, the guesser produces up to $\texttt{guess\_top\_k}$ guesses.

\bheading{Judger.}
Each PII instance is associated with $\texttt{guess\_top\_k}$ guesses. We select the top $\texttt{judge\_top\_k}$ guesses and pass them to a judger LLM, which labels each guess as \texttt{yes}, \texttt{no}, or \texttt{less precise}, corresponding to scores of 1.0, 0.0, and 0.5, respectively.

\bheading{Configuration.}
In our experiments, we set both $\texttt{guess\_top\_k}$ and $\texttt{judge\_top\_k} $ to 1. The prompt used for LLMs are unmodified following prior work~\cite{staab2023beyond}

\subsection{Impact of Imperfect Detection}
\begin{figure}[t]
    \centering
    \begin{minipage}{0.23\textwidth}
        \centering
        \includegraphics[width=\linewidth]{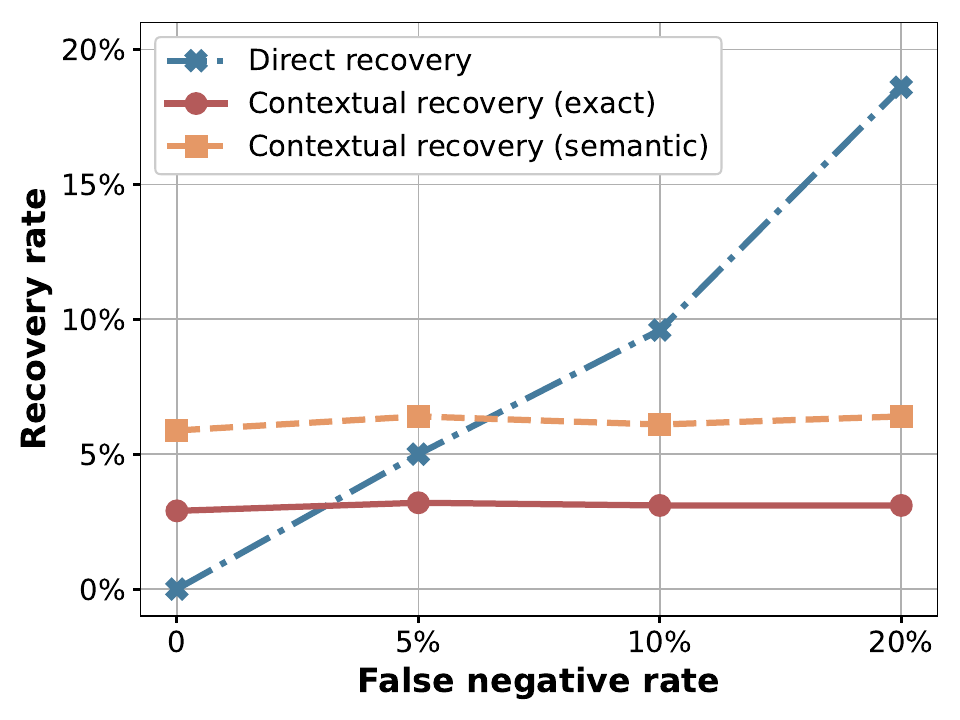}
        {\scriptsize (a) False negative}
        \label{fig:detecterror-a}
    \end{minipage}
    \hfill
    \begin{minipage}{0.23\textwidth}
        \centering
        \includegraphics[width=\linewidth]{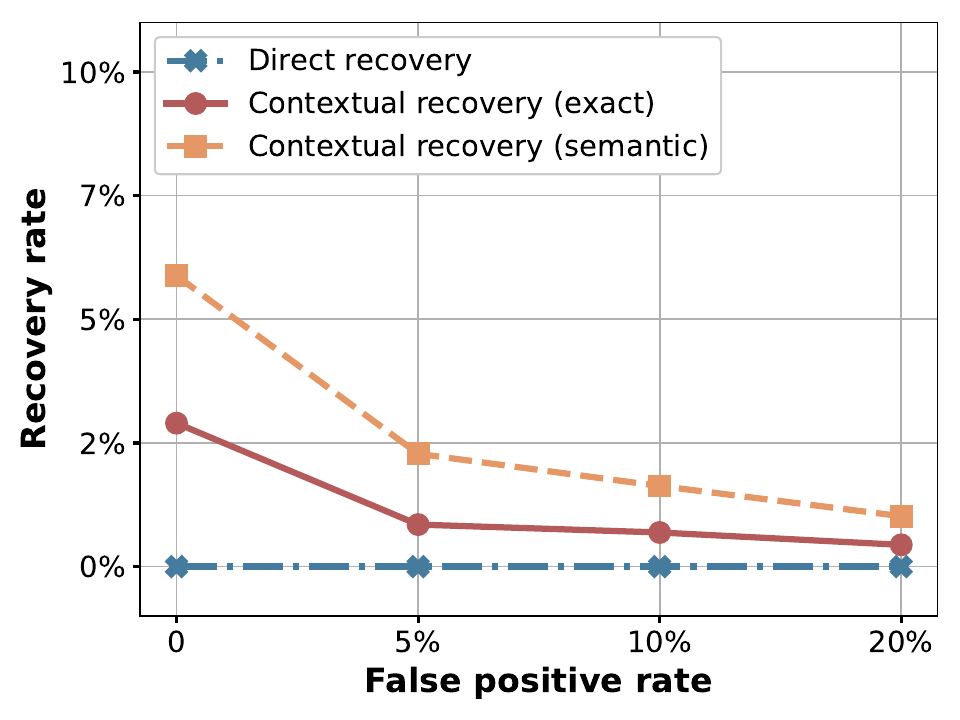}
        {\scriptsize (b) False positive}
        \label{fig:detecterror-b}
    \end{minipage}

    \caption{Impact of imperfect privacy detection (NarrativeQA).}
    \label{fig:detecterrornarra}
\end{figure}

\begin{figure}[t]
    \centering
    \begin{minipage}{0.23\textwidth}
        \centering
        \includegraphics[width=\linewidth]{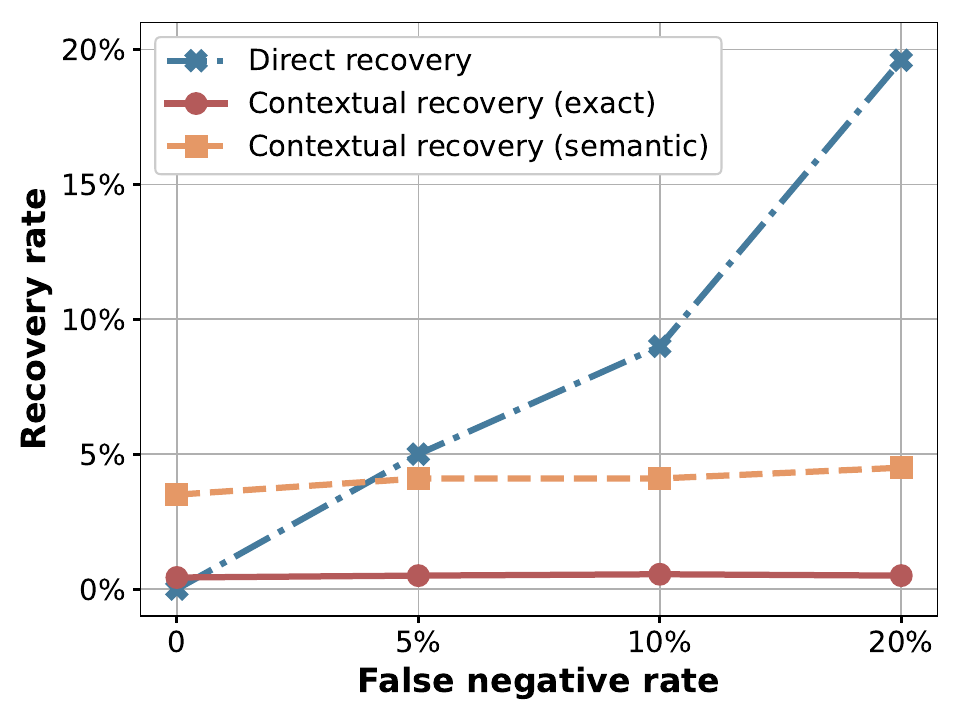}
        {\scriptsize (a) False negative}
        \label{fig:detecterror-a}
    \end{minipage}
    \hfill
    \begin{minipage}{0.23\textwidth}
        \centering
        \includegraphics[width=\linewidth]{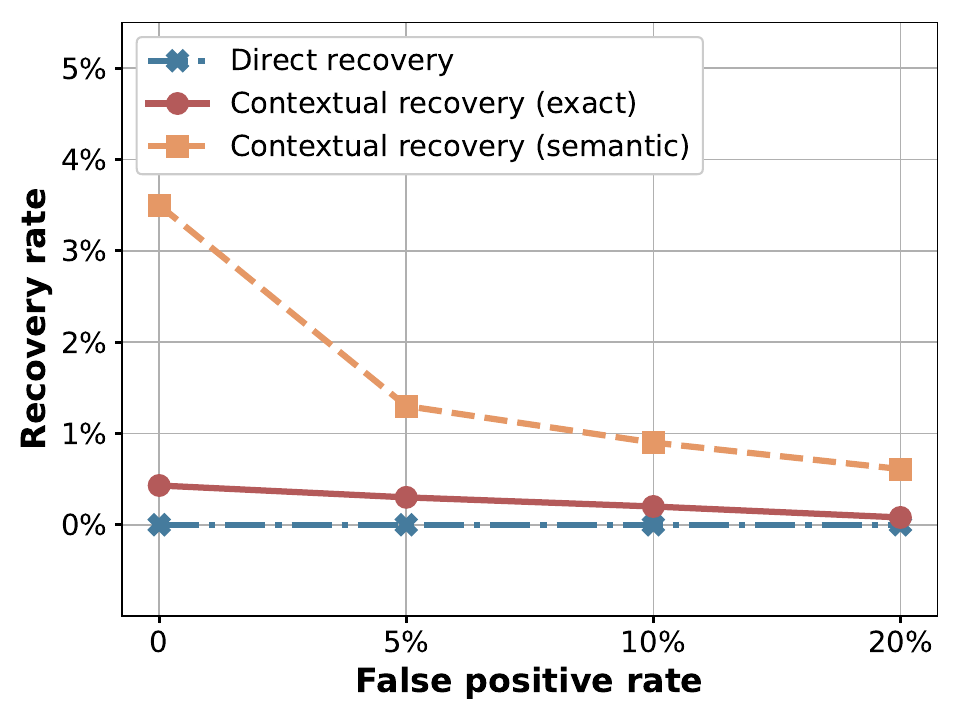}
        {\scriptsize (b) False positive}
        \label{fig:detecterror-b}
    \end{minipage}

    \caption{Impact of imperfect privacy detection (QMSum).}
    \label{fig:detecterrorqmsum}
\end{figure}

In addition to the results presented in the main paper on QASPER, we further evaluate \sysname on NarrativeQA (\figref{fig:detecterrornarra}) and QMSum (\figref{fig:detecterrorqmsum}) to examine the our findings. 
The results confirm the impact analyzed in the main paper.

\section{Generation Quality Evaluations}
\label{sec:appendix_generation_quality}

\begin{figure*}[!t]
\begin{minipage}{0.32\textwidth}
 	\centering
    \includegraphics[width=\textwidth]{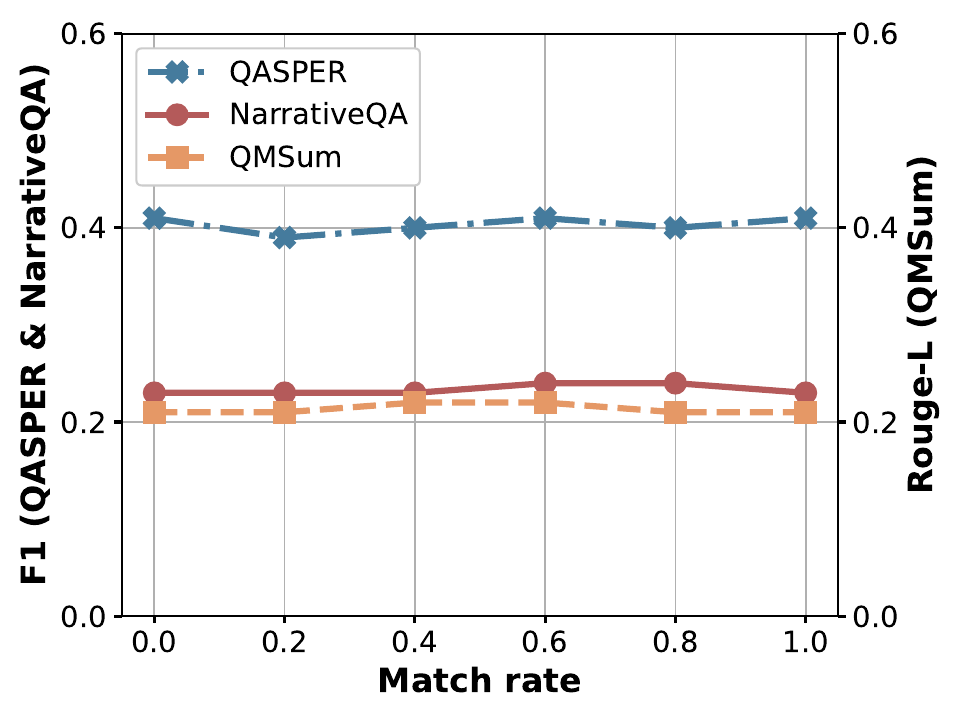}  
    \caption{Match rate impact (Qwen-7B).} \label{fig:matchrateimpactqwen7}
\end{minipage}
\begin{minipage}{0.1\textwidth}
\end{minipage}
\begin{minipage}{0.32\textwidth} 
    \includegraphics[width=\textwidth]{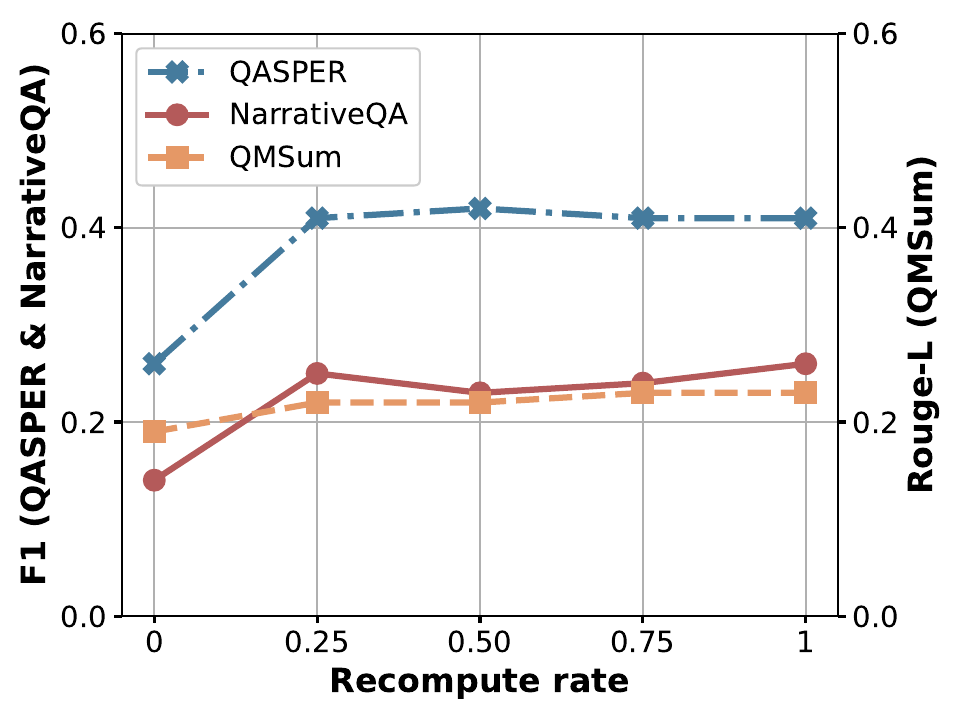}  
    \caption{Recompute rate impact (Qwen-7B).} \label{fig:recomputeimpactqwen7}
\end{minipage}
\begin{minipage}{0.1\textwidth}
\end{minipage}
\begin{minipage}{0.32\textwidth} 
    \includegraphics[width=\textwidth]{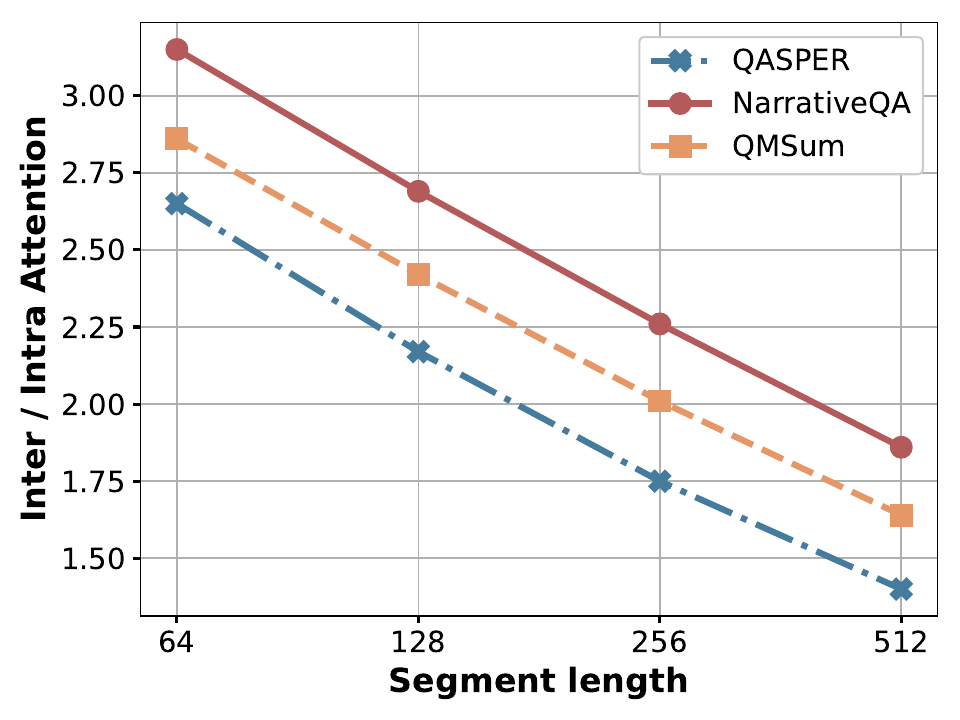}  
     \caption{Segment length impact (Qwen-7B).} \label{fig:segmentimpactqwen7}
\end{minipage}
\end{figure*} 

\begin{figure*}[!t]
\begin{minipage}{0.32\textwidth}
 	\centering
    \includegraphics[width=\textwidth]{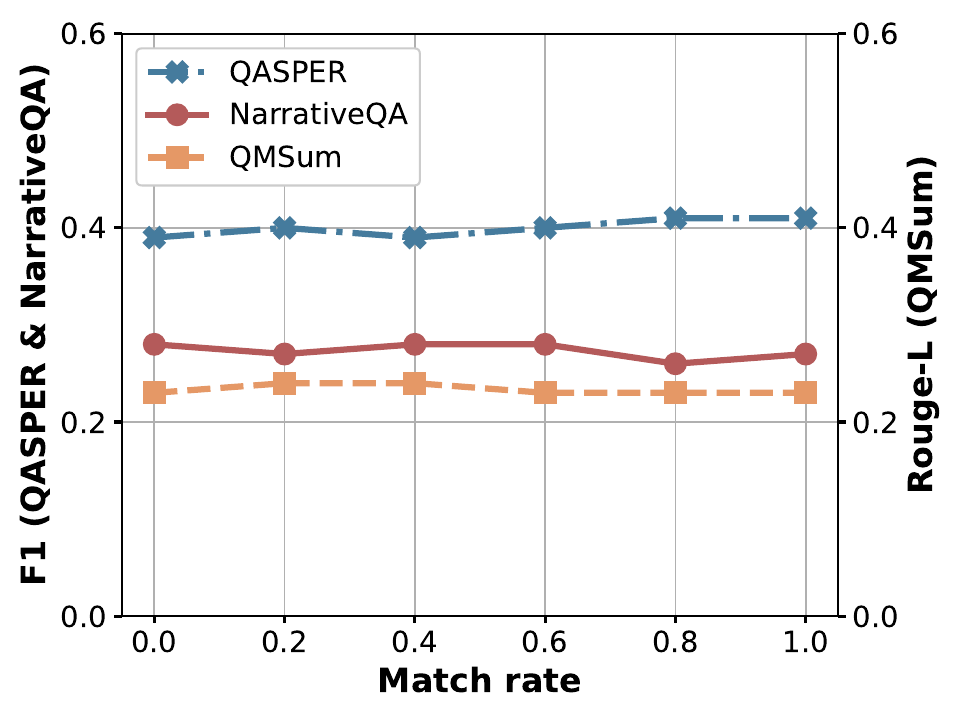}  
    \caption{Match rate impact (Qwen-14B).} \label{fig:matchrateimpactqwen14}
\end{minipage}
\begin{minipage}{0.1\textwidth}
\end{minipage}
\begin{minipage}{0.32\textwidth} 
    \includegraphics[width=\textwidth]{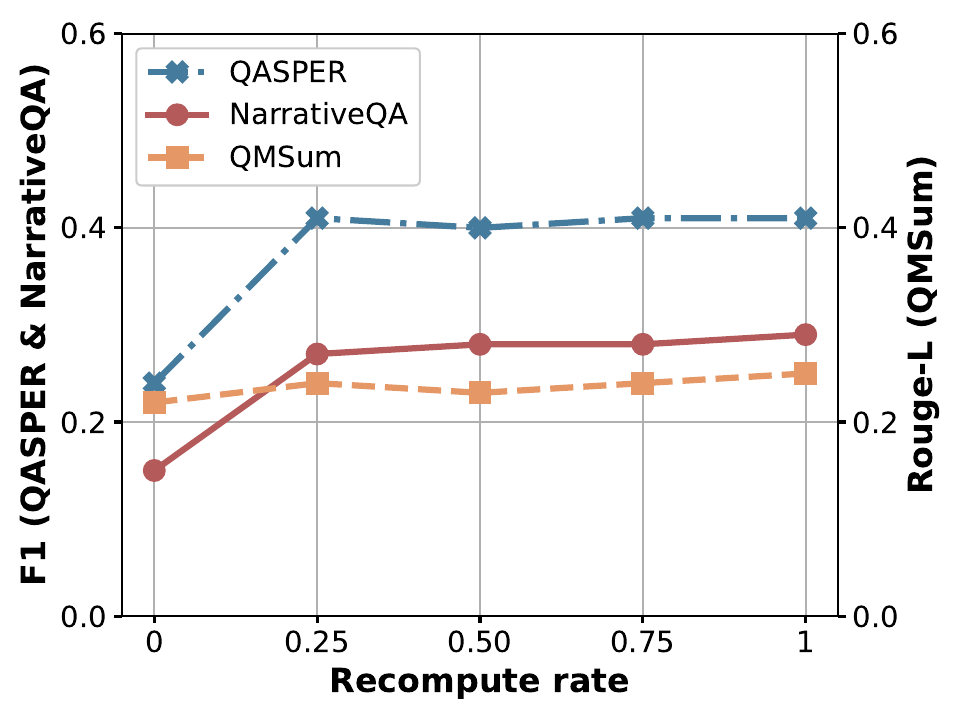}  
    \caption{Recompute rate impact (Qwen-14B).} \label{fig:recomputeimpactqwen14}
\end{minipage}
\begin{minipage}{0.1\textwidth}
\end{minipage}
\begin{minipage}{0.32\textwidth} 
    \includegraphics[width=\textwidth]{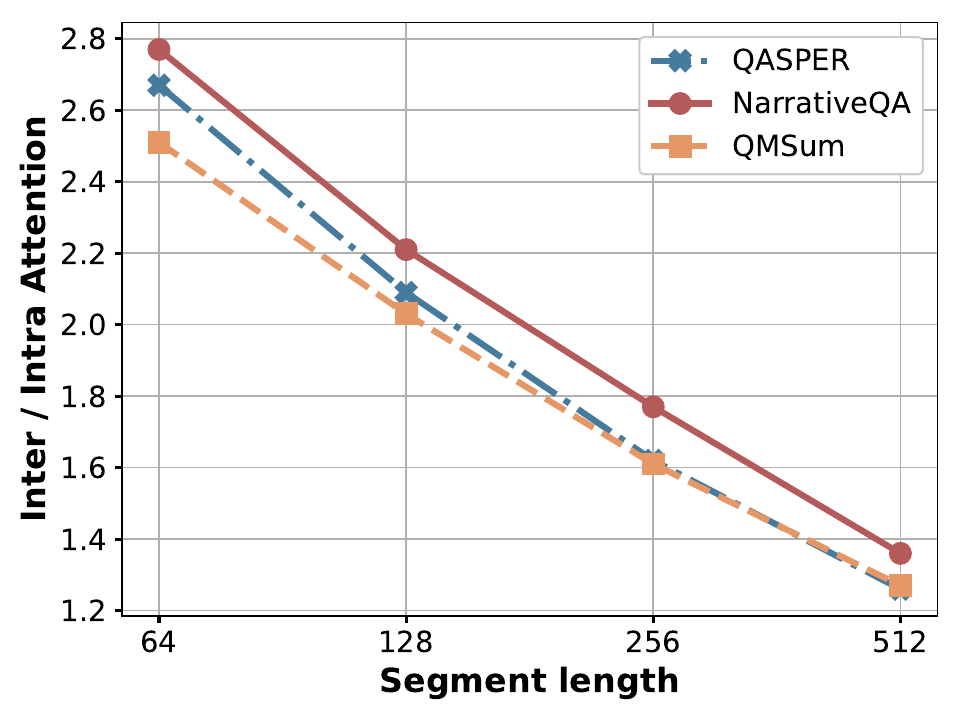}  
     \caption{Segment length impact (Qwen-14B).} \label{fig:segmentimpactqwen14}
\end{minipage}
\end{figure*}

In addition to the results presented in the main paper on Mistral-7B, we further evaluate \sysname on Qwen2.5-7B and Qwen2.5-14B to examine the generality of our findings. 
We follow the same experimental setup as described in Sec.~\ref{subsec:rq1}, including dataset construction and evaluation metrics. 
Overall, we observe consistent trends across models (\figref{fig:matchrateimpactqwen7} to \figref{fig:segmentimpactqwen14}): generation quality remains stable under varying match rates, improves with moderate recomputation, and benefits from longer segment lengths. 
These results confirm that the impact of the three key factors analyzed in Sec.~\ref{subsec:rq1} generalizes across different model architectures and scales.

\section{Efficiency Evaluations}
\label{sec:appendix_efficiency}
\begin{figure}[t]
    \centering
    \begin{minipage}{0.23\textwidth}
        \centering
        \includegraphics[width=\linewidth]{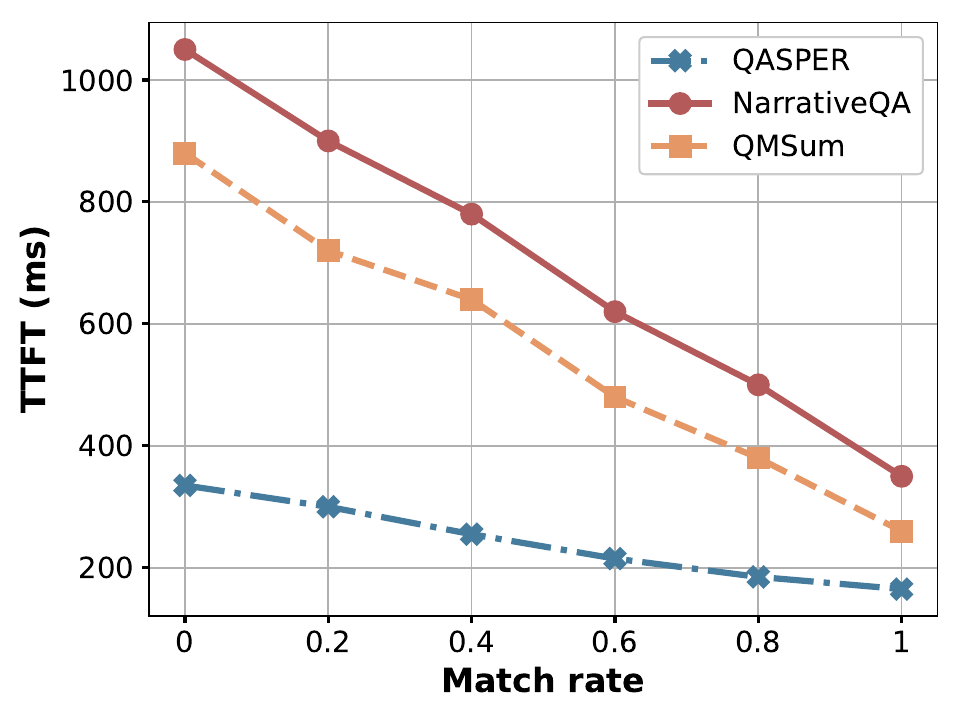}
        {\scriptsize (a) TTFT}
        \label{fig:match-a}
    \end{minipage}
    \hfill
    \begin{minipage}{0.23\textwidth}
        \centering
        \includegraphics[width=\linewidth]{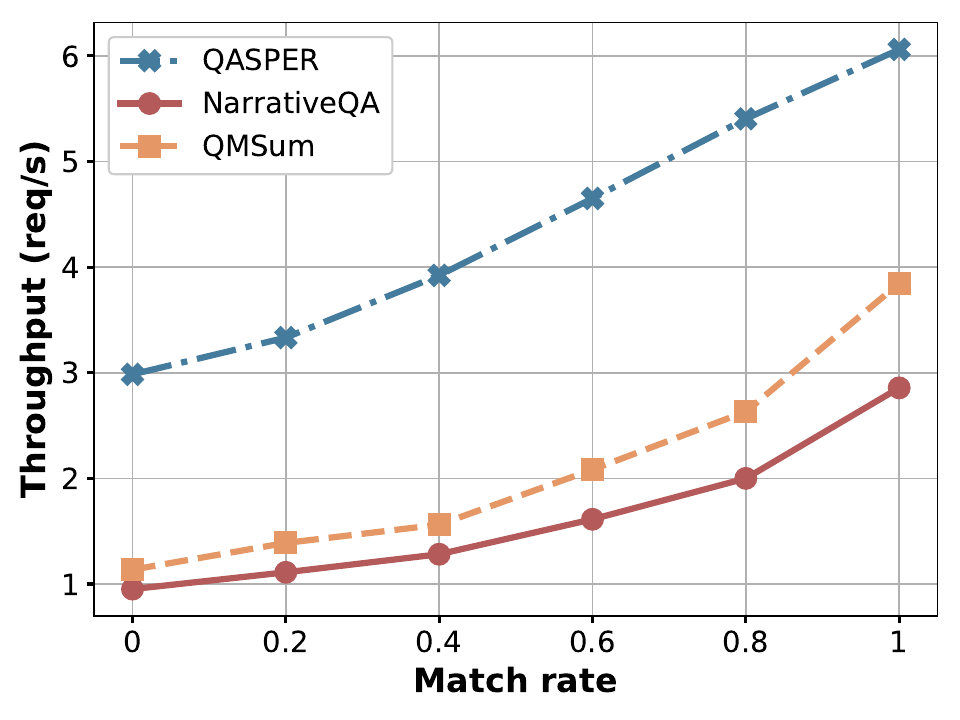}
        {\scriptsize (b) Throughput}
        \label{fig:match-b}
    \end{minipage}

    \caption{Impact of match rate on efficiency (Qwen-7B).}
    \label{fig:matchimpacteffqwen7}
\end{figure}

Similarly, we further evaluate \sysname on Qwen2.5-7B and Qwen2.5-14B to examine the generality of our findings. 
We follow the same experimental setup as described in Sec.~\ref{subsec:rq2}, including dataset construction and evaluation metrics. 
Overall, we observe consistent trends across models (from \figref{fig:matchimpacteffqwen7} to \figref{fig:recomputeimpacteffqwen14}), confirming that the efficiency of \sysname generalize across different model architectures and scales.

\begin{figure}[t]
    \centering
    \begin{minipage}{0.23\textwidth}
        \centering
        \includegraphics[width=\linewidth]{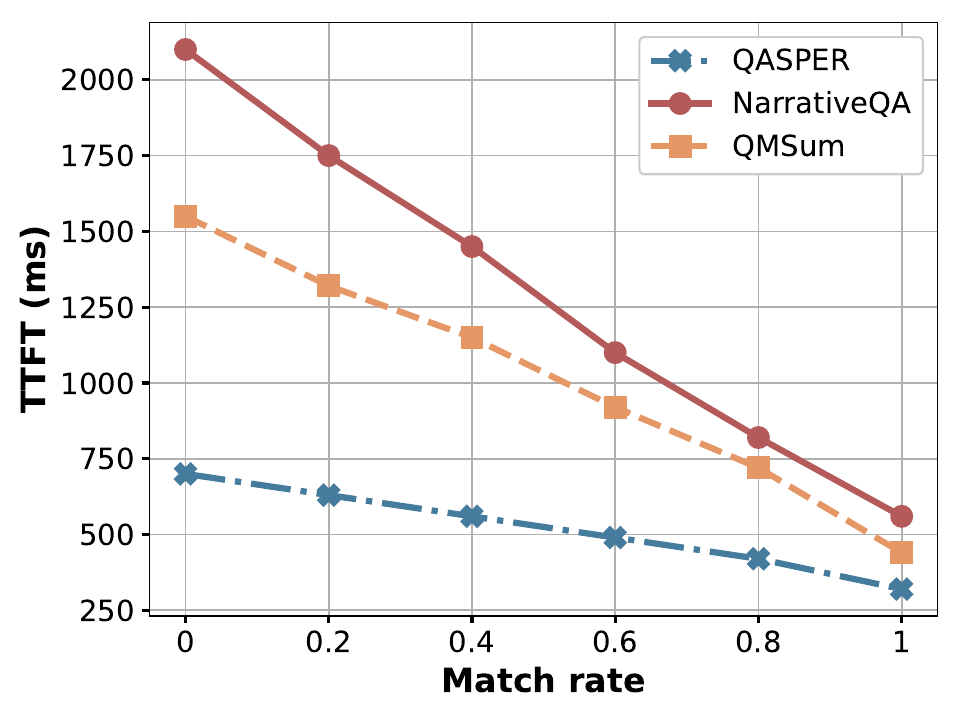}
        {\scriptsize (a) TTFT}
        \label{fig:match-a}
    \end{minipage}
    \hfill
    \begin{minipage}{0.23\textwidth}
        \centering
        \includegraphics[width=\linewidth]{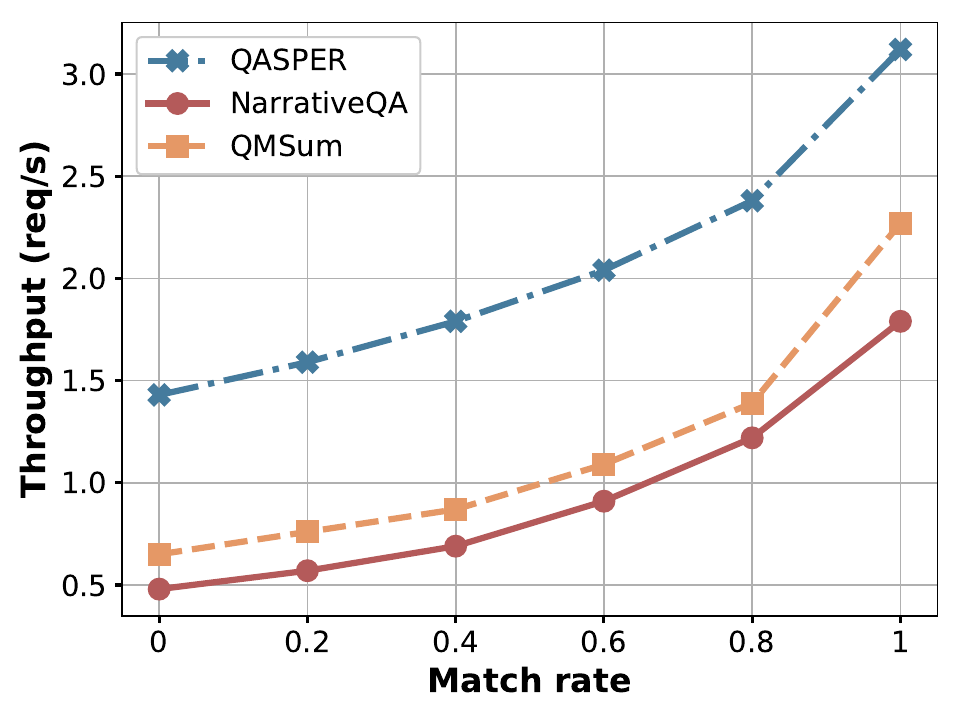}
        {\scriptsize (b) Throughput}
        \label{fig:match-b}
    \end{minipage}

    \caption{Impact of match rate on efficiency (Qwen-14B).}
    \label{fig:recomputeimpacteffqwen14}
\end{figure}

\begin{figure}[t]
    \centering
    \begin{minipage}{0.23\textwidth}
        \centering
        \includegraphics[width=\linewidth]{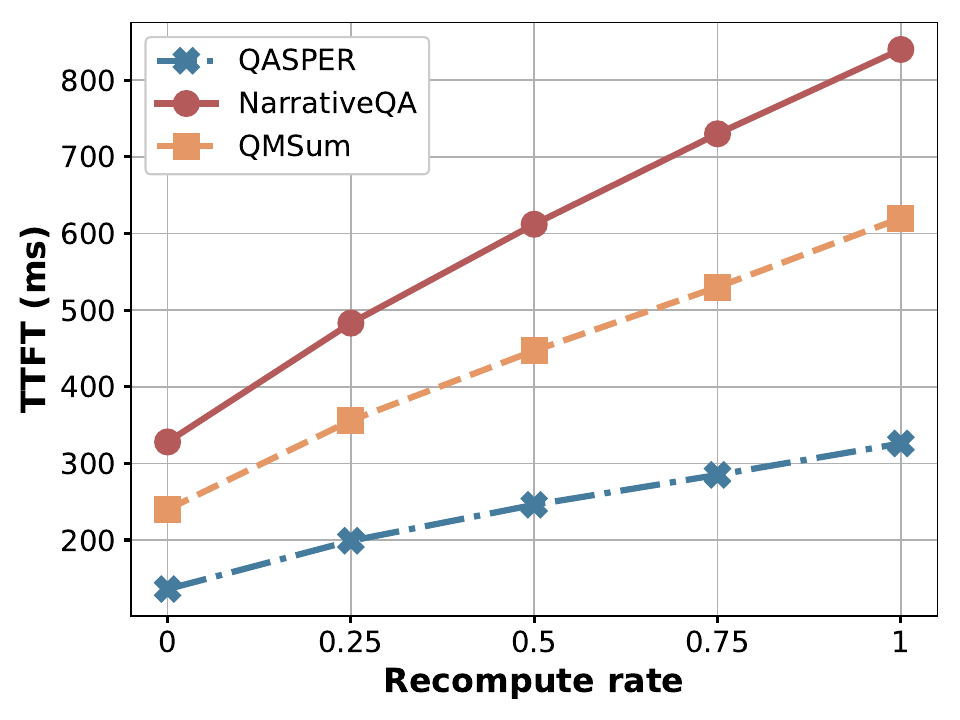}
        {\scriptsize (a) TTFT}
        \label{fig:match-a}
    \end{minipage}
    \hfill
    \begin{minipage}{0.23\textwidth}
        \centering
        \includegraphics[width=\linewidth]{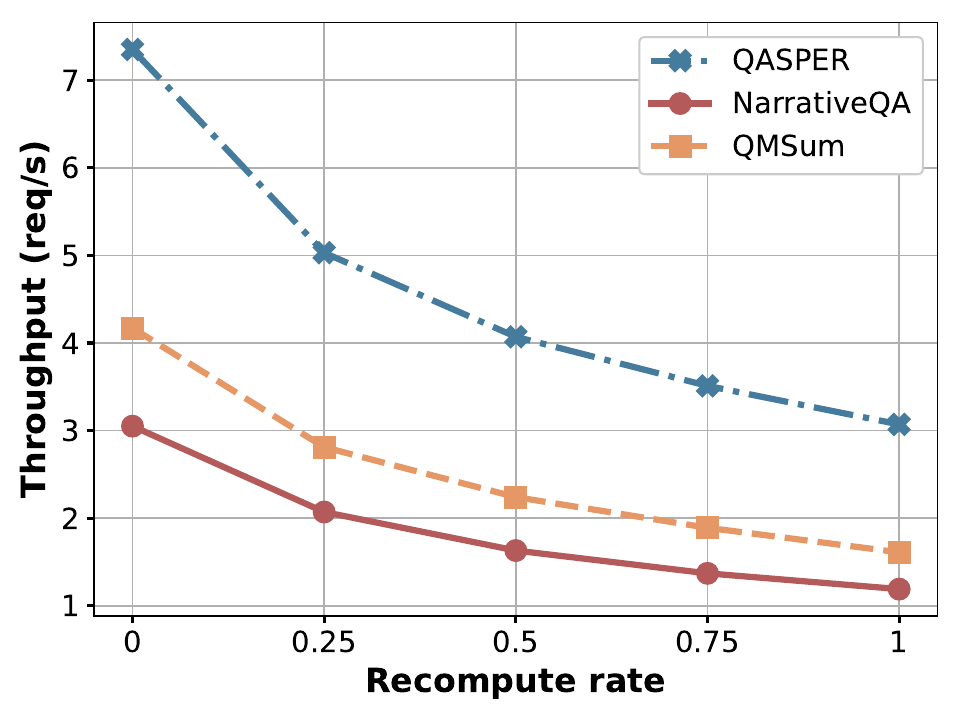}
        {\scriptsize (b) Throughput}
        \label{fig:match-b}
    \end{minipage}

    \caption{Impact of recompute rate on efficiency (Qwen-7B).}
    \label{fig:recomputeimpacteffqwen7}
\end{figure}

\begin{figure}[t]
    \centering
    \begin{minipage}{0.23\textwidth}
        \centering
        \includegraphics[width=\linewidth]{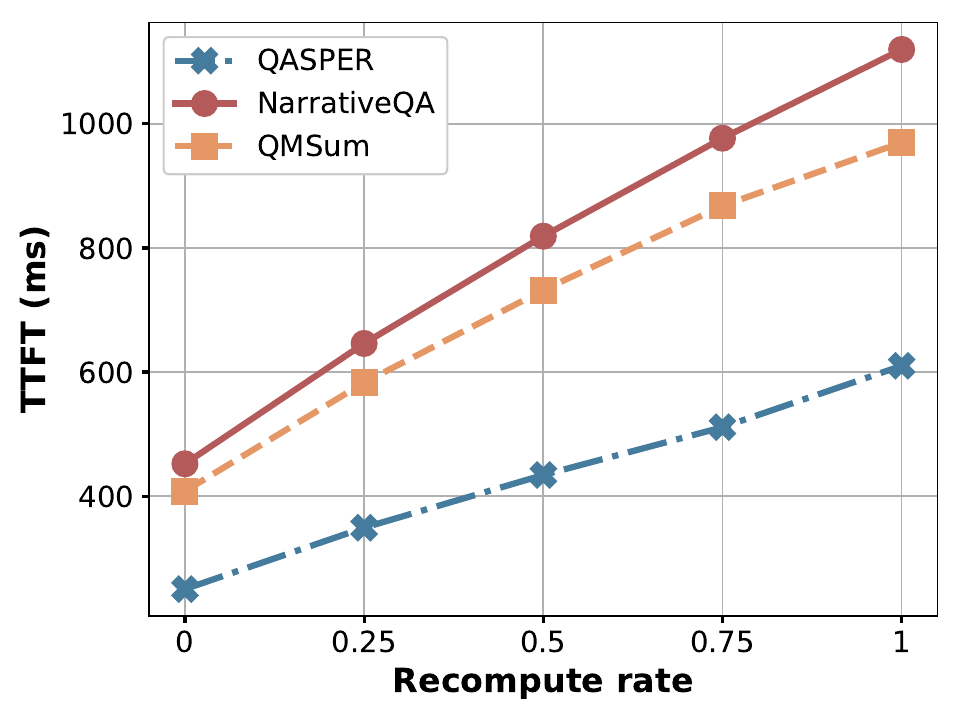}
        {\scriptsize (a) TTFT}
        \label{fig:match-a}
    \end{minipage}
    \hfill
    \begin{minipage}{0.23\textwidth}
        \centering
        \includegraphics[width=\linewidth]{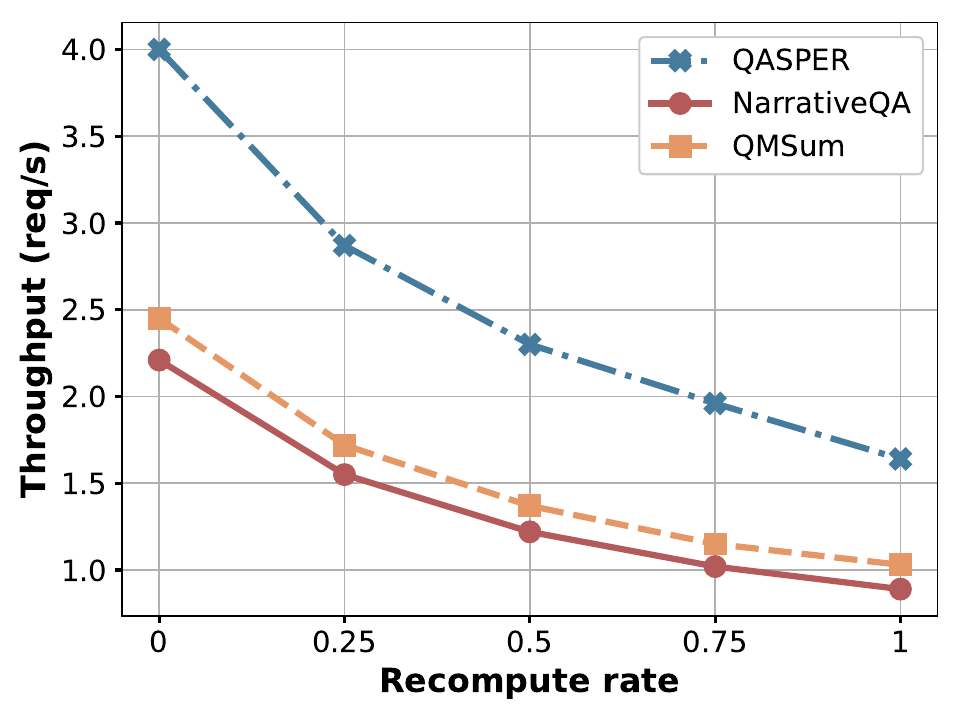}
        {\scriptsize (b) Throughput}
        \label{fig:match-b}
    \end{minipage}

    \caption{Impact of recompute rate on efficiency (Qwen-14B).}
    \label{fig:recomputeimpacteffqwen14}
\end{figure}

\begin{figure}[t]
    \centering
    \begin{minipage}{0.23\textwidth}
        \centering
        \includegraphics[width=\linewidth]{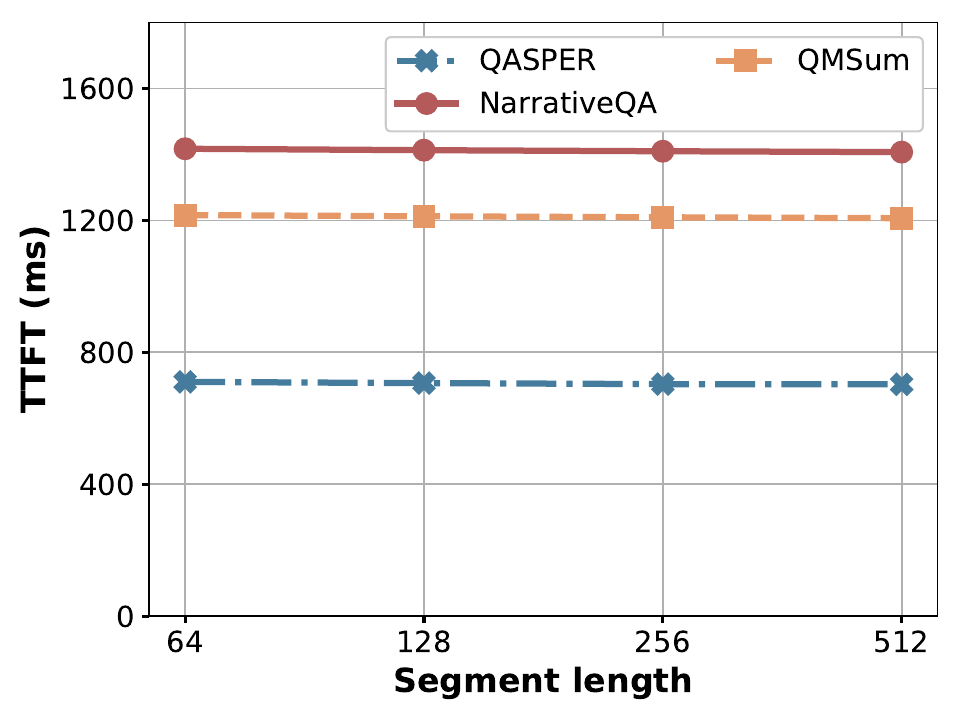}
        {\scriptsize (a) TTFT}
        \label{fig:length-a}
    \end{minipage}
    \hfill
    \begin{minipage}{0.23\textwidth}
        \centering
        \includegraphics[width=\linewidth]{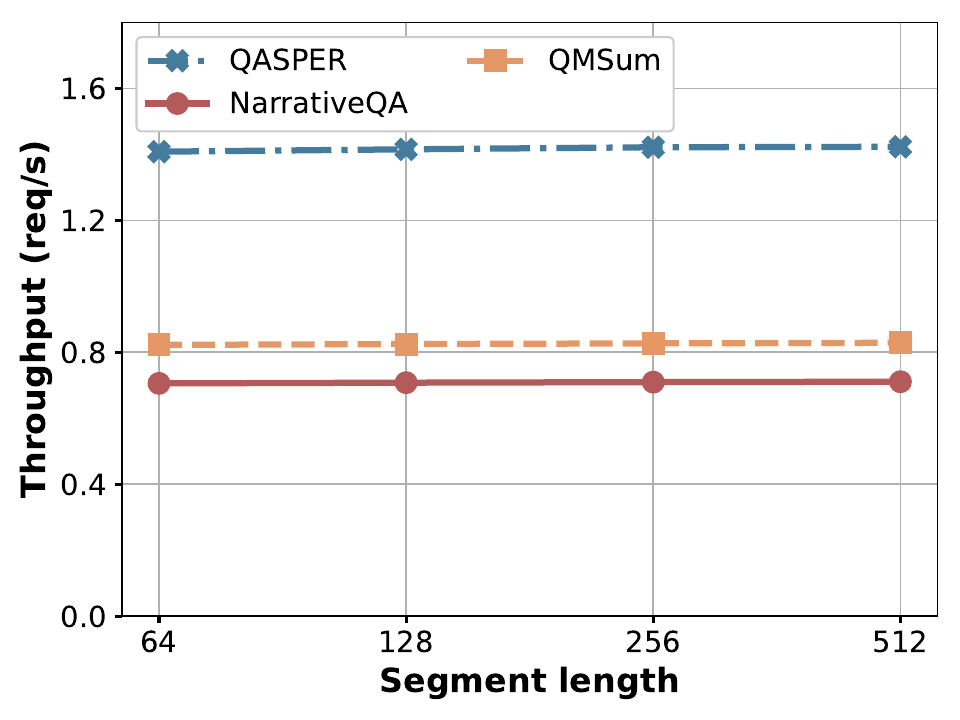}
        {\scriptsize (b) Throughput}
        \label{fig:length-b}
    \end{minipage}

    \caption{Impact of segment length on efficiency.}
    \label{fig:lengthimpacteff}
\end{figure}

\section{Privacy Detection Settings}
\label{appendix:privacy}

Following prior work on sensitivity identification in LLM prompts~\cite{ntwali2025detection}, we adopt Presidio~\cite{presidio}, a widely used privacy detection toolkit, as the default detector in our evaluation. Presidio supports multiple categories of sensitive entities, which we group into two classes:

\begin{packeditemize}
  \item \textbf{Personal identifiers}: CREDIT\_CARD, CRYPTO,IBAN\_CODE, EMAIL\_ADDRESS, NRP (passport), PERSON, PHONE\_NUMBER, SSN, US\_BANK\_NUMBER, US\_DRIVER\_LICENSE, US\_ITIN,US\_SSN, US\_PASSPORT.
  \item \textbf{Non-personal identifiers}: DATE\_TIME, IP\_ADDRESS, LOCATION, URL, AU\_ABN, AU\_ACN.
\end{packeditemize}

We configure three privacy levels for evaluation, as summarized in Table~\ref{tab:privacy-detector}.

\begin{table}[h]
\centering
\caption{Privacy detector configurations.}
\label{tab:privacy-detector}
\begin{tabular}{lp{6.2cm}}
\hline
\textbf{Level} & \textbf{Categories used} \\ \hline
Low & CREDIT\_CARD, CRYPTO, EMAIL\_ADDRESS, IBAN\_CODE, NRP, PERSON, PHONE\_NUMBER, SSN \\ \hline
Medium & All Personal identifiers \\  \hline
High & All Personal + Non-personal identifiers \\ \hline
\end{tabular}
\end{table}

This configuration allows us to emulate settings from lightweight filtering (Low) to strict protection of all detectable categories (High).

\begin{figure}[t]
    \centering
    \begin{minipage}{0.23\textwidth}
        \centering
        \includegraphics[width=\linewidth]{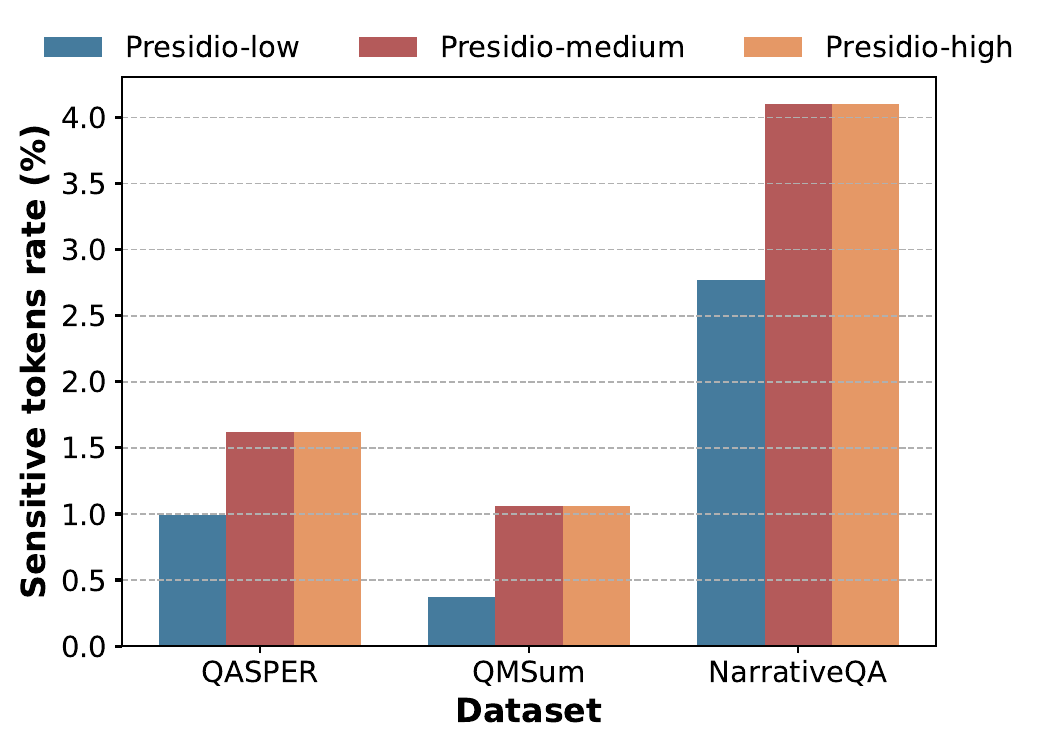}
        {\scriptsize (a) Match rate}
        \label{fig:length-a}
    \end{minipage}
    \hfill
    \begin{minipage}{0.23\textwidth}
        \centering
        \includegraphics[width=\linewidth]{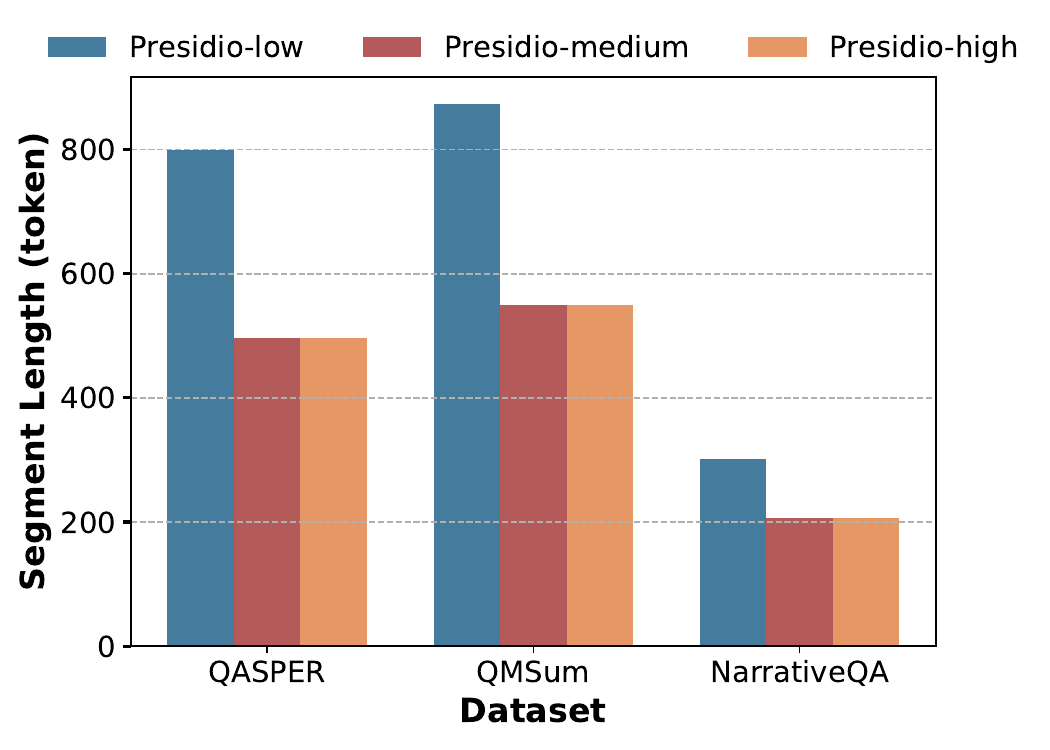}
        {\scriptsize (b) Segment length}
        \label{fig:length-b}
    \end{minipage}

    \caption{Impact of privacy detection methods.}
    \label{fig:privacyimpacteff}
\end{figure}

\end{document}